\documentclass[journal=jacsat,manuscript=article]{achemso}
\usepackage[version=3]{mhchem} 
\usepackage{color}
\author{Sandip Aryal}
\affiliation{Theoretical Division, Los Alamos National Laboratory, Los Alamos, New Mexico 87545, United States}
\author{Gaoxue Wang}
\affiliation{Theoretical Division, Los Alamos National Laboratory, Los Alamos, New Mexico 87545, United States}
\author{Enrique R. Batista}
\affiliation{Theoretical Division, Los Alamos National Laboratory, Los Alamos, New Mexico 87545, United States}
\email{erb@lanl.gov}

\title[An \textsf{achemso} demo]
 {Stoichiometric and Non-stoichiometric Cesium Potassium Antimonide Photocathodes: Ab-initio Insights into its Properties for Photoemission}

\begin{document}
\begin{abstract}
Alkali-metal antimonides, especially cesium-potassium-antimonide ($\rm CsK_2Sb$), are strong candidates for next-generation photocathodes in linear accelerators due to their low work-function, fast response, high quantum yield, and ability to operate under visible light. In this study, first-principles methods are used to examine the structural, electronic, optical, and surface properties of $\rm CsK_2Sb$ relevant to its photoemission performance. Our results for $\rm CsK_2Sb$ show strong absorption in the visible range consistent with experimental observations. The computed work-functions for the stable surfaces are significantly lower than the commonly used metallic  photocathodes. This material exhibits electron and hole mobilities of 111.86 $\rm cm^2/Vs$ and 3.24 $\rm cm^2/Vs$, respectively. Since real materials inherently contain intrinsic defects, we analyze native point defects in $\rm CsK_2Sb $ and identified Cs and K vacancies as most likely. These defects introduce mid-gap states, modify the absorption spectrum, and may significantly influence photoemission behavior. This work provides valuable insights into optimizing $\rm CsK_2Sb$ for high-efficiency photocathode applications.

\end{abstract}

\flushbottom
\maketitle

\thispagestyle{empty}


\section*{1. Introduction}
Photocathode-based electron beams are essential for applications such as free-electron lasers, energy recovery linacs, and ultrafast electron diffraction. \cite{T038, sommer1993, S581, S9324, SXG2023, X154, O935, G157, VTES2007, AAVA2007, GBBF2002, LJID2011} 
An ideal photocathode material should exhibit fast-response time, minimal thermal emittance, low work-function, and a wide spectral range for high quantum yield. Additionally, superior air stability is desirable to ensure a long operational lifetime. However, none of the currently used photocathode materials possess all the characteristics listed. Commonly used photocathodes can be classified into two main categories: (i) metals, \cite{Q1218, T2116, D0619, D0917, Q1020} and (ii) semiconductors. \cite{sommer1993, MGFJ2007, DDD1977, C1412, A8113, M2411, O0414, S0115, K9510, G203, C152} Whereas metallic photocathodes offer great stability against contamination from residual gases, giving them long lifetimes, and display fast response times, their large work-function (4.65 eV for Cu, 3.66 eV for Mg, and 4.25 eV for Pb) \cite{M1997} forces the operators to drive them using ultraviolet light. \cite{GMAM2017} Additionally, a small electron escape depth of only a few atomic layers\cite{MGFJ2007} combined with enhanced electron-electron scattering and high reflectivity from metal surfaces, leads to a very low quantum yield of  ($10^{-4}-10^{-1}$) for this class of photocathodes, \cite{SXG2023} which limits their ability to generate the intense beams needed for linear accelerators.

On the other hand, semiconductor photocathode materials \cite{PPCW2023, V956, DBJA2013, GK2019, PGAJ2022, C152, WRB2014, LJID2011, DSK1995, G17_2017, Wang2025, Aryal2025} have their own pros and cons. While more reactive and sensitive to residual gases,\cite{SXG2023, vitaly_2016} semiconducting photocathode materials have lower work-function facilitating the photoemission process, a long mean free path with an escape depth of 10 to 20 atomic layers, \cite{MGFJ2007} and broad spectral sensitivity ranging from the visible to ultraviolet region. These qualities contribute to their high quantum yield, in excess of 10\%. \cite{SXG2023} As a result, semiconductor photocathodes are emerging as strong candidates to replace metallic photocathodes as high-brightness electron sources for linear accelerators in the near future.

Alkali-metal antimonides \cite{PPCW2023, V956, DBJA2013, GK2019, PGAJ2022, C152, WRB2014, LJID2011, DSK1995} are among the most promising semiconductor photocathode materials for next-generation electron emitters in linear accelerators. One of the most extensively studied alkali-metal antimonide photocathodes is $\rm CsK_2Sb$. \cite{GK2019, V956, WRB2014, ILAB2011, LJID2011, DKKE1993, LBF2010, GMN2023, CSMJ2019, LBKD2010, XKML2022}
This material exhibits performance similar to that of cesium antimonide ($\rm Cs_3Sb$), another promising photocathode material from the same family, yet, its reduced thermionic emission at room temperature offers a distinct advantage in low-photocurrent scenarios, where the thermionic emission in $\rm Cs_3Sb$ limits its performance. \cite{sommer1993}  
The high quantum efficiency of $\rm CsK_2Sb$, approximately 30\% when excited with a 400 nm laser\cite{sommer1993, MGFJ2007} and around (8–12)\% for a (527–532) nm laser, \cite{WRB2014, LJID2011, DSK1995} pave the way for its application in generating high-brightness electron beams for next-generation linear accelerators.
While $\rm CsK_2Sb$ degrades even in the high vacuum conditions of the accelerator tube due to remaining oxygen, water, and hydrocarbons, with an operational lifetime of typically 3–30 hours,  \cite{DKKE1993, DSK1995, WRB2014} mitigation methodologies to improve their stability and inertness have been proposed and are an active area of research.\cite{G17_2017, WYNB2018, GLYHYM_2020} Therefore, the potential of $\rm CsK_2Sb$ as a photocathode material remains promising.

The standard framework for modeling photoemission in materials is Spicer's three-step model, \cite{S9324, S581} which divides the process into three stages: (i) optical absorption, (ii) carrier transport, and (iii) carrier escape from the surface. For optimal photoemission efficiency, the material must have a high optical absorption coefficient to effectively absorb photons, thereby exciting a greater number of electrons from the valence to the conduction band. Additionally, high carrier mobility and electrical conductivity will enable the transport of excited electrons to the surface with minimal scattering. Lastly, a low work-function is desirable, as it facilitates carrier escape from the surface. Thus, a comprehensive understanding of the optical, electrical transport, and surface properties of $\rm CsK_2Sb$ is crucial for gaining deeper insight into its photoemission behavior.

 Photocathode crystals grown in the laboratory inevitably contain a finite concentration of defects at a finite temperature, which can and do influence each stage of the photoemission process. These defects can  introduce electron and/or hole traps, altering the optical, transport, and surface properties of the material—key factors in determining its photoemission. Therefore, a first-principles comprehensive study of the physics of these materials will include insight into the formation of native point defects, and the impact of each kind of defect on the electronic and optical properties of the material. Such study is essential for optimizing its photoemission and identify which type of defect should be prioritized for mitigation. Experimentally, non-stoichiometric \(\rm Cs_{3+x}Sb\), another widely used photocathode material from the same alkali-metal-antimonide family, has been synthesized within the range \(\rm -0.05 \leq x \leq 0.04\), \cite{JCMB1989}. This deviation of the stoichiometric \(\rm Cs_3Sb\) arises due to the presence of defects. Motivated by this, we extend our study to non-stoichiometric $\rm CsK_2Sb$, where native point defects such as vacancies, antisites, and interstitials contribute to compositional deviations.  
This study specifically investigates non-stoichiometric $\rm CsK_2Sb$ while ensuring that the \(N_{\frac{Cs+K}{Sb}}\) ratio—representing the total number of Cs and K atoms relative to Sb—remains within the range \(3.04 \geq N_{\frac{Cs+K}{Sb}} \geq 2.95\).

In this work, we leveraged density functional theory (DFT) to investigate the electronic, elastic, vibrational, electrical transport, and surface properties of $\rm CsK_2Sb$ that govern its photoemission.  
Our calculations predict a small bandgap of 1.13 eV for $\rm CsK_2Sb$, indicating its sensitivity to visible light. The ordered $\rm CsK_2Sb$ is found to be dynamically stable, in contrast to \(\rm Cs_3Sb\) and \(\rm K_3Sb\), where the \(\rm DO_3\) symmetry-broken distorted phase is more stable.\cite{Aryal2025, nangoi2022} $\rm CsK_2Sb$ exhibits low elastic moduli, suggesting weak interatomic interactions and high susceptibility to deformation. The computed room-temperature electron and hole mobilities for $\rm CsK_2Sb$ are 111.86 \(\rm cm^2/Vs\) and 3.24 \(\rm cm^2/Vs\), respectively. Calculations show that Cs and K vacancies are the preferred point defects in $\rm CsK_2Sb$ compared to Sb vacancies, antisites, and interstitials of Cs, K, and Sb. These vacancy defects introduce states near the band edges of $\rm CsK_2Sb$, contributing to low-energy peaks in its absorption spectrum. The stability of the $\rm CsK_2Sb$ surface depends on the chemical potential of its constituents. For example, the (111)-Cs surface is stable under Cs-rich conditions, regardless of \(\rm \mu_{K}\). As Cs-deficient regime is approached,  the (111)-K(b) terminated surface becomes stable over a wide range of \(\rm \mu_{Cs}\), independent of \(\rm \mu_{K}\).  
The stable (111)-Cs and (111)-K(b) surfaces exhibit the work-function of approximately 2.14 eV and 2.56 eV, which are comparable to the work-function of 2.01 eV for the semiconducting \(\rm Cs_3Sb\)  photocathodes,\cite{WYNB2018} and significantly lower than that of metallic photocathodes: \(\sim\)4.65 eV for Cu, \(\sim\)3.66 eV for Mg, and \(\sim\)4.25 eV for Pb. \cite{M1997} The strong absorption of $\rm CsK_2Sb$ in the visible region, combined with its high carrier mobility and small work-function, suggests that this material is a promising alternative to metallic photocathodes.

This paper is organized as follows. Section 2 outlines the computational methods. Section 3 presents our results on the electronic, vibrational, elastic, electrical transport, and surface properties of $\rm CsK_2Sb$. Additionally, it includes the stability, electronic, and optical properties of non-stoichiometric $\rm CsK_2Sb$. Finally, Section 4 summarizes our findings.
\section*{2. Computational Methods}
\subsection*{2.1 Bulk calculations}
Plane-wave periodic density functional theory (DFT), as implemented in the Vienna Ab initio Simulation Package (VASP), \cite{KF1996PRB, KF1996CMS} was used for all calculations in this work. The Perdew-Burke-Ernzerhof (PBE) functional \cite{PBE1996} was employed to approximate the exchange-correlation energy, while the projected augmented wave (PAW) method \cite{B1994, KJ1999} was used to account for interactions between valence electrons and ions.
To account for van der Waals dispersion, the DFT-D3 correction by Grimme et al. \cite{G2006} was included. It may be noted that the DFT-D3 correction is found to yield a lattice constant in good agreement with the experimental value for $\rm CsK_2Sb$. For geometry optimization of the defect supercell, a Monkhorst-Pack \cite{MP1776} $k$-mesh of \(2\times2\times2\) was employed, while a denser \(5\times5\times5\) $k$-point grid was used for density of states (DOS) calculations. The defect supercells were relaxed until the residual forces on each atom were below \(5\times10^{-3}\) eV/{\AA}. Energy convergence thresholds were set at \(10^{-6}\) eV for structural relaxations and \(10^{-9}\) eV for optical property calculations. A kinetic energy cutoff of 400 eV was applied throughout, which we found to be sufficently large to reach converged results. Phonon band structures were computed using VASP \cite{KF1996PRB, KF1996CMS} in conjunction with the Phonopy package\cite{TT2015, TT2023}.

Previous work on $\rm CsK_2Sb$ has shown that excitonic effects in this material are minimal, indicating that its optical spectra can be accurately described using the independent particle approximation.\cite{CC_2021}
The optical properties of both pristine and defective $\rm CsK_2Sb$ were calculated using VASP within the independent particle approximation. In this approximation, the imaginary part of the complex dielectric function, $\epsilon(\omega)$, is given by \cite{G0644}
\begin{equation}
 \rm \epsilon_{imag}^{\alpha\beta}(\omega) = \frac{4\pi^2e^2}{\Omega} \lim_{q \to 0}\frac{1}{q^2}\sum_{c,v,k}2w_k\delta(\epsilon_{ck}-\epsilon_{vk}-\omega)<u_{ck+e_{\alpha}q}|u_{vk}><u_{ck+e_{\beta}q}|u_{vk}>^{*}
\label{EO1}
\end{equation}
In Equation \ref{EO1}, $\rm e_{\alpha}$ represents the unit vector along the Cartesian directions, while $\rm w_k$ and $\rm \Omega$ denote the weight of the k-point and the volume of the unit cell, respectively. The indices c, v, and k correspond to conduction band states, valence band states, and k-points. Using the Kramers-Kronig relation, the real part of the complex dielectric function can be derived from the imaginary part as follows \cite{G0644}:
\begin{equation}
\rm \epsilon_{real}^{\alpha\beta}(\omega) = 1 + \frac{2}{\pi}P \int_0^{\infty}\frac{ \epsilon_{imag}^{\alpha\beta}(\omega^{\prime})\omega^{\prime}}{\omega^{\prime 2} - \omega^2}d\omega^{\prime}. 
\label{EO2}
\end{equation}
In the Equation \ref{EO2}, P denotes the principal value of the integral. The real  and imaginary part  of the complex refractive index can be expressed as $\rm n(\omega) = \sqrt{\frac{|\epsilon(\omega)| + \epsilon_{real}(\omega)}{2}}$ and $\rm \kappa(\omega) = \sqrt{\frac{|\epsilon(\omega)| - \epsilon_{real}(\omega)}{2}}$, respectively, where  $|\epsilon(\omega)|$ represents the absolute value of the complex dielectric function. The absorption coefficient is given by $\rm \alpha = 4\pi\kappa/\lambda$.  
\subsection*{2.2 Surface calculations}
The properties of low Miller indices surfaces—such as-(001), (110), and (111)-were investigated as these, in general, represent the most stable and naturally occurring facets of a crystal. These surfaces were constructed from the conventional unit cell of $\text{CsK}_2\text{Sb}$, with various surface terminations of its constituents systematically examined in this study. To model the surfaces of $\rm CsK_2Sb$, a slab-type configuration was used, imposing periodic bondary conditions along the $x$- and $y$-directions while a large vacuum region ($\geq 20$~\AA) was introduced along the transverse $z$-direction to minimize interactions between periodic images along this direction.
The lattice constants of the (001) slabs are $a = b = 8.56$~\AA~in the periodic $x$- and $y$-directions. Similarly, the (110) slab has lattice parameters of $a = 12.11$~\AA~and $b = 8.56$~\AA~along the $x$- and $y$-directions, respectively, while (111) slabs have lattice constants of $a = b = 12.11$~\AA.

All atoms within 5–6~\AA~of the surface of the slabs were relaxed while keeping the middle layer fixed at the bulk configuration. The optimization was considered converged when the residual force on each atom was less than 0.005 eV/\AA. The energy convergence criterion was set to $10^{-6}$ eV for both bulk structural relaxations and surface property calculations. To mitigate potential electrostatic artifacts introduced by the presence of defects and surface terminations, a dipole correction was applied during slab relaxation.
Monkhorst-Pack \cite{MP1776} $k$-grids of $3\times3\times1$, $2\times3\times1$, and $3\times3\times1$ were used for relaxation of the (001), (110), and (111) slabs, respectively. Denser Monkhorst-Pack \cite{MP1776} $k$-meshes of $7\times7\times1$, $7\times9\times1$, and $7\times7\times1$ were employed to compute the DOS.  
The slab thicknesses considered in this study range from 2 to 3 nm. It may be noted that previous work on a similar material, $\text{Cs}_3\text{Sb}$, has shown that a slab thickness of approximately 1.5 nm is sufficient to achieve converged surface properties such as the work-function \cite{WYNB2018}.
\section*{3. Results and Discussion}
According to Spicer,\cite{S9324,S581} the photoemission in a material depends on the material's ability to absorb light, their carrier transport properties, and the escape rate of carriers from its surface. In this section, a comprehensive understanding of the stability, optical, transport and surface properties of crystalline $\rm CsK_2Sb$ have been presented. Since defects form inevitably in materials and are expected to affect photoemission, formation of native point defects have also been explored. This section is structured as follows: Subsection 3.1 discusses the electronic and vibrational properties of $\rm CsK_2Sb$. Subsection 3.2 explores the elastic properties of $\rm CsK_2Sb$.
Subsection 3.3 discusses the carrier transport properties of $\rm CsK_2Sb$. 
Subsection 3.4 examines the  stability of defects in $\rm CsK_2Sb$ , and its influence on the electronic, and optical properties of $\rm CsK_2Sb$. Finally subsection 3.5 discusses the surface properties of $\rm CsK_2Sb$ including surface stability, DOS, and work-function / ionization potential. 
\subsection*{3.1 Electronic and Vibrational Properties of $\rm CsK_2Sb$}

\begin{figure}[ht!]
\centering
\includegraphics[width=10cm]{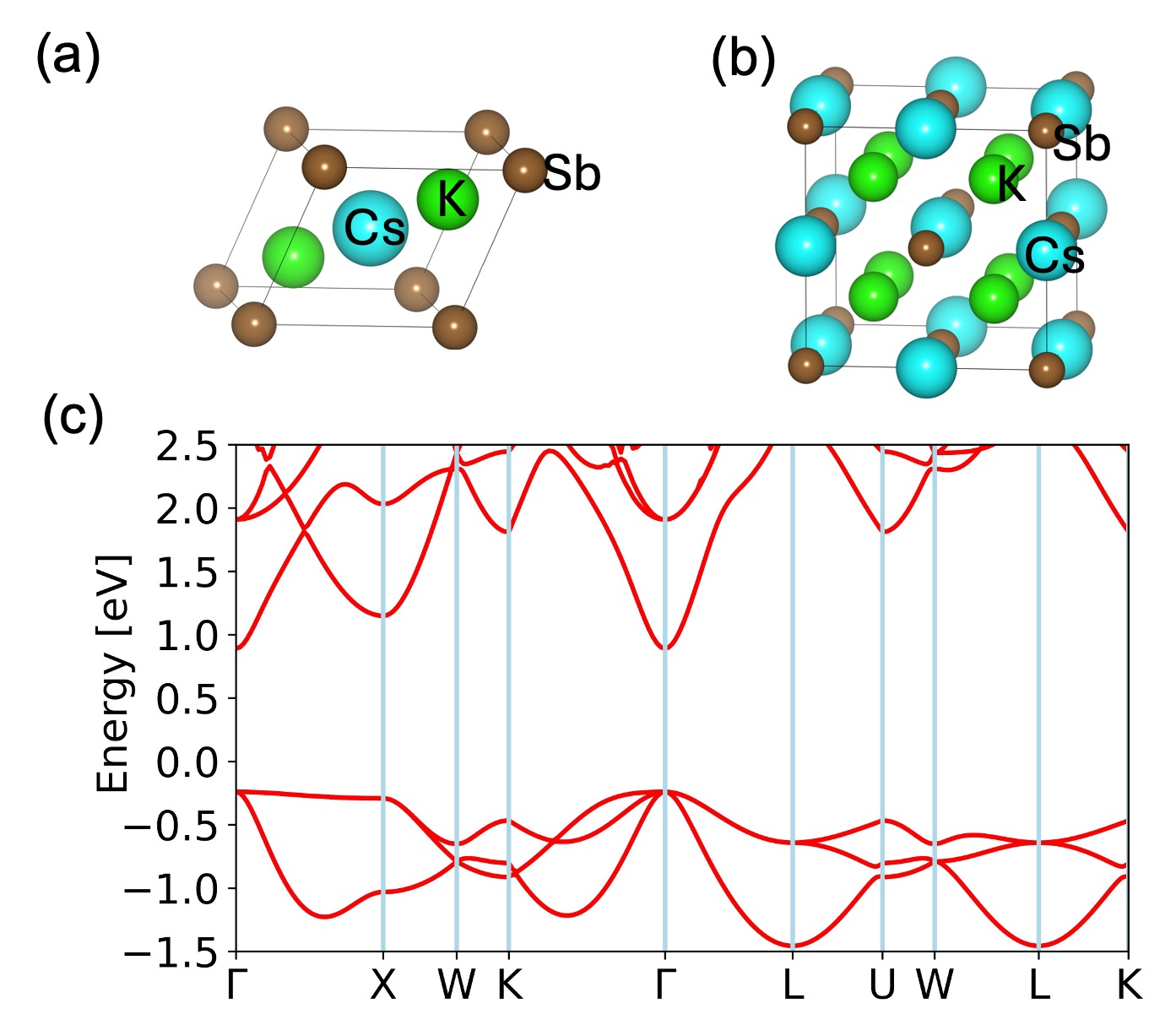}
\caption{(a) Primitive, (b) conventional unit cells, and (c) Electronic band structure of $\rm CsK_2Sb$ obtained at PBE+D3 level, showing a direct bandgap of 1.13 eV at the $\Gamma$-point. }
\label{fig:Fig1a}
\end{figure}

We begin by examining the structural and electronic properties of bulk $\rm CsK_2Sb$, which crystallizes in a face-centered cubic $\rm DO_3$ phase with space group $Fm\bar{3}m$, as illustrated in Figure \ref{fig:Fig1a}(a-b). In this structure, Cs cations occupy the Wyckoff position (1/2, 1/2, 1/2), while K cations are positioned at (1/4, 1/4, 1/4), and Sb anions reside at (0,0,0). Additionally, each Cs cation is coordinated to eight K cations as its nearest neighbors, with a bond distance of 3.71\AA{} (PBE+D3 level). Similarly, each K cation is bonded to four Cs cations and four Sb anions, all at an equal bond distance of 3.71\AA{} (PBE+D3 level). The lattice constants of $\rm CsK_2Sb$ computed at different levels of approximations are summarized in Table~\ref{tab:table1}.  The calculated equilibrium lattice constant of 8.56~\AA~at the PBE+D3 level is within 0.6\% of the experimental lattice constant of 8.615~\AA~obtained from x-ray diffraction analyses.\cite{DDD1977}  As shown in Table~\ref{tab:table1}, the lattice constant values obtained employing the hybrid (HSE06) and meta-GGA functionals (SCAN and $r^2$SCAN) are also in excellent agreement with experimental data, further supporting the reliability of these methods for structural predictions.

\begin{table}[ht!]
\caption{\label{tab:table1} Comparison of the computed lattice constant (a) and bandgap ($\rm E_g$) for $\rm Cs K_2 Sb$ with experimental results.}
\centering
\begin{tabular}{c c c c} 
\hline\hline
$ \rm Method  $ & a ({\AA}) & $\rm E_g (eV)$\\
\hline
$\rm PBE   $  & 8.76 & \rm 0.85 \\ 
$\rm PBE+D3 $     & 8.56 & \rm 1.13 \\
$\rm HSE06 $     & 8.67 & \rm 1.59 \\
$\rm SCAN $     & 8.65 & \rm 1.54 \\
$\rm r^2SCAN $     & 8.63 & \rm 1.57 \\
$\rm Experiment   $ & \rm 8.615 \cite{DDD1977}& 1.20\cite{GV1978} \\
\hline\hline
\end{tabular}
\end{table}

\begin{figure}[ht!]
\centering
\includegraphics[width=10cm]{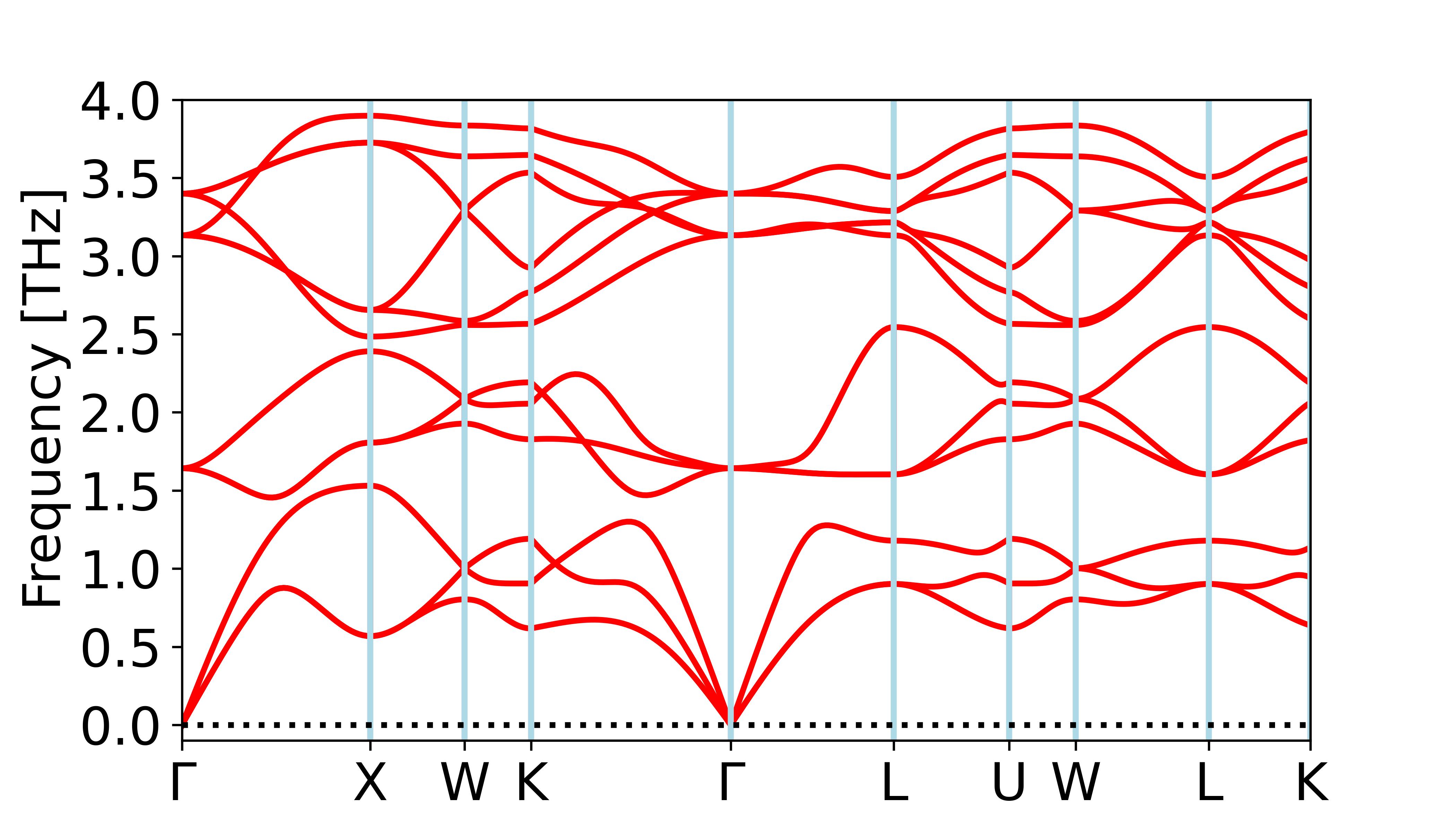}
\caption{Phonon dispersion for $\rm CsK_2Sb$ showing no imaginary modes indicating its dynamical stability.}
\label{fig:Fig1b}
\end{figure}

While the PBE functional is known to underestimate the electronic bandgap of $\rm CsK_2Sb$, the inclusion of dispersion corrections via PBE+D3 yields a direct bandgap of 1.13 eV at the ${\Gamma}$-point (Figure~\ref{fig:Fig1a}(c)), in close agreement with the experimental value of 1.20 eV obtained from photoconductivity measurements.\cite{GV1978} To evaluate the reliability of different exchange-correlation functionals, like the lattice constant, we also computed the bandgap using the hybrid (HSE06) and the meta-GGA (SCAN and $r^2$SCAN) functionals. The corresponding bandgap values are summarized in Table~\ref{tab:table1}. Our results indicate that both SCAN and $r^2$SCAN predict bandgaps comparable to HSE06, albeit slightly larger than those obtained from PBE+D3 and experimental photoconductivity measurements in Ref. \citenum{GV1978}.

The stability of $\rm CsK_2Sb$ was examined by calculating its phonon dispersion. Figure \ref{fig:Fig1b} shows no imaginary modes in the phonon spectrum, indicating its  dynamical stability. The phonon dispersion was obtained using the primitive unit cell shown in Figure \ref{fig:Fig1a}(a) and consists of 12 modes in total. The three low-frequency modes are acoustic, while the remaining high-frequency modes are optical, the highest at 3.90 THz.
Unlike the ordered $\rm CsK_2Sb$, the ordered $\rm K_3Sb$ (see Figure {\color{blue}{S1}}) and ordered $\rm Cs_3Sb$ \cite{nangoi2022, Aryal2025} exhibit imaginary modes in their phonon dispersion, indicating their dynamical instability. However, the $\rm DO_3$ symmetry-broken distorted $\rm K_3Sb$ (Figure {\color{blue}{S1}}) and distorted $\rm Cs_3Sb$ \cite{nangoi2022, Aryal2025} exhibit no imaginary modes in their phonon dispersion, indicating that these distorted phases are dynamically more stable than their ordered counterparts. Therefore, in this work, the elastic and carrier transport properties of ordered $\rm CsK_2Sb$ are compared with those of the stable distorted phases of $\rm Cs_3Sb$ and $\rm K_3Sb$. 
\subsection*{3.2 Elastic Properties of $\rm CsK_2Sb$}
Besides ensuring that the phonon spectrum does not have modes with imaginary frequencies, the stability of the crystal structure should  also be evaluated by analyzing the elastic tensor and Born stability criteria. Due to their cubic unit cell structures, these materials exhibit three independent elastic constants: $\rm C_{11}$, $\rm C_{12}$, and $\rm C_{44}$. The computed elastic constants, presented in Table \ref{tab:table1}, satisfy the Born stability criteria for cubic crystals, expressed as \cite{M1446}
\begin{align}
    C_{11} - C_{12} &> 0, \nonumber \\
    C_{11} + 2C_{12} &> 0, \nonumber \\
    C_{44} &> 0.
\end{align}
Applying these conditions to the computed elastic constants (Table \ref{tab:table1}) shows that the DFT-predicted structure is mechanically stable.  
The computed elastic constants were then used to determine the Voigt and Reuss bounds for the cubic crystals and extract macroscopic elastic properties, Young's modulus, and Poisson's ratio. The Voigt bounds are given by \cite{H5247}
\begin{align}
    K_V &= \frac{1}{3} (C_{11} + 2C_{12}); \quad G_V = \frac{1}{5} (C_{11} - C_{12} + 3C_{44})
\end{align}
Likewise, the Reuss bounds for these materials are given by \cite{H5247}
\begin{align}
    K_R &= \frac{1}{3} (C_{11} + 2C_{12}); \quad G_R = \frac{5(C_{11} - C_{12})C_{44}}{4C_{44} + 3(C_{11} - C_{12})}
\end{align}
The Voigt-Reuss-Hill (VRH) average was then computed from the Voigt and Reuss bounds using \cite{H5247}
\begin{align}
    K &= \frac{K_V + K_R}{2}; \quad G = \frac{G_V + G_R}{2},
\end{align}
Here, $K$ and $G$ represent the bulk and shear moduli of elasticity, respectively.  
The Young’s modulus ($E$) and Poisson’s ratio ($\nu$) are derived from $K$ and $G$ using \cite{H5247}  
\begin{align}
    E &= \frac{9KG}{3K+G}; \quad \nu = \frac{3K - 2G}{2(3K + G)}
\end{align}
\begin{table}[ht]
\caption{\label{tab:table1} Calculated elastic constants ($C_{11}$, $C_{12}$, and $C_{44}$), bulk modulus ($K$), shear modulus ($G$), Young’s modulus ($E$), and Poisson’s ratio ($\nu$) for $\rm CsK_2Sb$, distorted $\rm Cs_3Sb$, and distorted $\rm K_3Sb$.}
\centering
\begin{tabular}{c c c c c c c c} 
\hline\hline
 & $\rm C_{11}$ & $\rm C_{12}$ & $\rm C_{44}$ & $\rm K$ & $\rm G$ & $\rm E$ & $\rm \nu$  \\
&  (GPa) & (GPa) & (GPa) & (GPa) & (GPa) & (GPa) &  \\
\hline
$\rm CsK_2Sb$ & 19.35 & 8.97 & 9.59 & 12.43 & 7.50 & 18.73 & 0.25\\
$\rm Cs_3Sb$\cite{Aryal2025} & 11.12 & 6.69 & 6.11 & 8.17 & 4.07 & 10.47 & 0.29\\
$\rm K_3Sb$ & 13.83 & 9.70 & 10.74 & 11.08 & 5.64 & 14.47 & 0.28\\
\hline\hline
\end{tabular}
\end{table}

The calculated moduli of elasticity for $\rm CsK_2Sb$ agree well with the reported values from an earlier study. \cite{LBF2010}  For the distorted $\rm Cs_3Sb$ and distorted $\rm K_3Sb$, no prior work on their elastic properties is available for direct comparison. However, Table \ref{tab:table1} shows that they exhibit elastic constants and moduli of elasticity comparable to those of $\rm CsK_2Sb$.  It may be noted that the values for distorted $\rm Cs_3Sb$ have been taken from our earlier work (see Ref. \citenum{Aryal2025}) for direct comparison.
Table \ref{tab:table1} also reveals that the elastic moduli of these materials are significantly lower than those of Cu—another widely used metallic photocathode material, for which $K = 140.2$ GPa, $G = 45.4$ GPa, $E = 123.5$ GPa, and $\nu = 0.35$ at room temperature. \cite{L7448} 

It is worth noting that elastic moduli determine a material’s resistance to deformation and are closely related to the strength of chemical bonds. Low elastic moduli values indicate relatively weak chemical bonds in these materials, which is consistent with their large K-Sb, Cs-Sb, and Cs-K bond lengths, suggesting weaker interatomic interactions and lower resistance to structural distortions. Additionally, the elastic moduli provide valuable insight into whether a material is likely to be ductile or brittle. Pugh’s criterion \cite{T01} suggests that a material is ductile when the ratio $K/G > 1.75$, whereas it is brittle if the ratio falls below this threshold.  
Based on this criterion, we predict that $\rm CsK_2Sb$ is brittle, whereas distorted $\rm Cs_3Sb$ and $\rm K_3Sb$ are ductile. Furthermore, Pugh's criterion establishes a threshold value of $\nu = 0.26$ to distinguish between ductile and brittle behavior. Since $\nu < 0.26$ for $\rm CsK_2Sb$, it is classified as brittle. Conversely, as $\nu > 0.26$ for distorted $\rm Cs_3Sb$ and distorted $\rm K_3Sb$, these materials are identified as ductile.

The elastic constants presented in Table \ref{tab:table1} were used to compute the electrical transport properties of these alkali-metal-antimonide photocathode materials, which will be discussed in the next section.
\subsection*{3.3 Electrical Transport Properties of $\rm CsK_2Sb$}
Photoemission in a material depends not only on carrier transport properties, particularly conductivity and mobility, but also on carrier lifetime. A longer electron and hole lifetime, combined with higher mobility and conductivity, is desirable for an efficient photocathode material, as it ensures that carriers can rapidly reach the surface with minimal scattering after optical excitation and before being ejected.  
In this work, the scattering rates, conductivity, and mobility of electrons and holes were computed using the AMSET package.\cite{GPFW2021} AMSET calculates transport properties by applying the relaxation time approximation (RTA) to solve the Boltzmann transport equation. The code incorporates multiple scattering mechanisms to describe how electrons and holes interact with phonons and impurities, including (a) acoustic deformation potential (ADP) scattering, (b) ionized impurity (IMP) scattering, and (c) polar optical phonon (POP) scattering. The total scattering rate ($\frac{1}{\tau}$) was obtained by summing the contributions from these individual scattering mechanisms (Matthiessen’s rule) as \cite{GPFW2021}
\begin{equation}
\rm \frac{1}{\tau}  =  \frac{1}{\tau_{ADP}} +  \frac{1}{\tau_{IMP}} +  \frac{1}{\tau_{POP}}
\end{equation}
The relaxation times, denoted as \( \tau_{\rm ADP} \), \( \tau_{\rm IMP} \), and \( \tau_{\rm POP} \), correspond to the ADP, IMP, and POP scattering mechanisms, respectively. The electronic relaxation time (\(\tau\)) is given by the inverse of the total scattering rate in a scattering process.

Among these mechanisms, POP scattering is intrinsically inelastic, meaning it involves the absorption or emission of phonons, leading to a change in the electron's energy. In contrast, ADP and IMP scattering are elastic processes, ensuring that the electron's energy remains unchanged.  
The electronic transitions from an initial state \( |n\textbf{k} \rangle \) to a final state \( |m\textbf{k}+\textbf{q} \rangle \), for both elastic and inelastic processes, are calculated using Fermi’s Golden Rule. The inelastic scattering rate is given by: \cite{GPFW2021}
\begin{align}
\tau^{-1}_{n\textbf{k} \to m\textbf{k}+\textbf{\textbf{q}}} &= \frac{2\pi}{\hbar} \left| g_{nm}(\textbf{k},\textbf{q}) \right|^2  
\times\big[ (n_\textbf{q} +1 - f^{0}_{m\textbf{k}+\textbf{\textbf{q}}})\delta(\Delta \epsilon^{nm}_{\textbf{k},\textbf{q}} - \hbar\omega_\textbf{q})  \nonumber \\
&\quad + (n_\textbf{q} + f^{0}_{m\textbf{k}+\textbf{\textbf{q}}})\delta(\Delta \epsilon^{nm}_{\textbf{k},\textbf{q}} + \hbar\omega_\textbf{q}) \big]
\end{align}
The elastic scattering rate is:\cite{GPFW2021}
\begin{equation}
\tau^{-1}_{n\textbf{k} \to m\textbf{k}+\textbf{\textbf{q}}} = \frac{2\pi}{\hbar} \left|g_{nm}(\textbf{k},\textbf{q})\right|^2 \delta(\epsilon_{n\textbf{k}} - \epsilon_{m\textbf{k}+\textbf{\textbf{q}}})
\end{equation}
where, \( \Delta \epsilon^{n,m}_{\textbf{k},\textbf{q}} = \epsilon_{n\textbf{k}} -\epsilon_{m\textbf{k}+\textbf{\textbf{q}}} \), $n_\textbf{q}$ is the Bose-Einstein distribution function, $f^{0}_{n\textbf{k}}$ is the Fermi-Dirac distribution function, and \( \delta \) represents the Dirac delta function.
 \( g_{nm}(\textbf{k},\textbf{q}) \) is the scattering matrix element, which describes the probability amplitude for scattering  from an initial electronic state \( |n\textbf{k} \rangle \) to a final state \( |m\textbf{k}+\textbf{q} \rangle \) via a phonon with wave vector $\textbf{q}$ and frequency $\omega_\textbf{q}$.\cite{GPFW2021} This matrix element can be expressed as follows:\cite{GPFW2021}
\begin{equation}
g_{nm}(\textbf{k},\textbf{q}) =  \langle m\textbf{k}+\textbf{\textbf{q}} | \Delta_\textbf{q}V | n\textbf{k} \rangle 
\end{equation}
where \( n \) represents the band index,   \( \textbf{k} \) is the wave vector, \( \hbar \) is the reduced Planck’s constant, and \( \Delta_\textbf{q}V \) is the electronic perturbation.

Similar to the total scattering rate, the total mobility (\(\mu\)) follows Matthiessen’s rule, which expresses it as a combination of the mobility components limited by POP, IMP, and ADP scattering as:
\begin{equation}
\frac{1}{\mu}  =  \frac{1}{\mu_{\rm ADP}} +  \frac{1}{\mu_{\rm IMP}} + \frac{1}{\mu_{\rm POP}}
\end{equation}
where \( \mu_{\rm ADP} \), \( \mu_{\rm IMP} \), and \( \mu_{\rm POP} \) represent the mobility contributions limited by ADP, IMP, and POP scattering, respectively.

The material properties used as inputs for the AMSET package are presented in Table S1. In Table S1, the effective polar phonon frequency ($\omega_{po}$) is calculated as a weighted sum over all the $\Gamma$-point phonon modes. $\omega_{po}$ is then used in computing the contribution of polar-optical phonon scattering.\cite{GPFW2021} Since the PBE functional tends to underestimate the bandgap in most semiconductors, the bandgaps of $\rm CsK_2Sb$, distorted $\rm Cs_3Sb$, and distorted $\rm K_3Sb$ were evaluated using the more accurate hybrid functional (HSE06) \cite{HS2004,HSE2003} at the PBE-relaxed lattice parameter. The HSE bandgap at the PBE lattice parameter was subsequently used to compute the electrical transport properties of these alkali-metal-antimonides.
\begin{figure}[ht!]
\centering
\includegraphics[width=19cm]{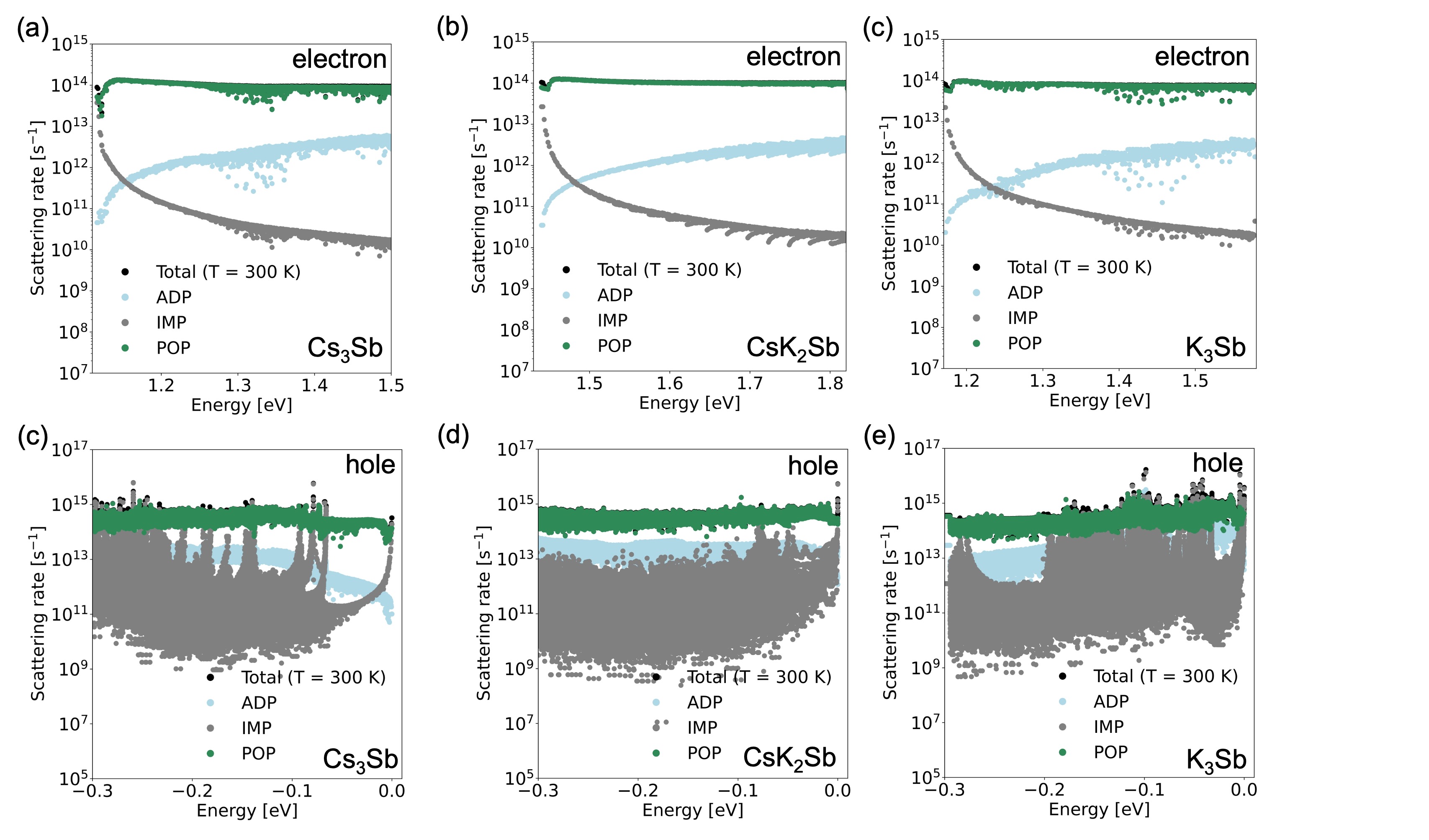}
\caption{ADP, IMP, and POP contributions to (a–c) electron scattering rates and (d–f) hole scattering rates for $\rm Cs_3Sb$, $\rm CsK_2Sb$, and $\rm K_3Sb$ at $\rm T = 300$ K for an electron and hole concentration of $10^{16}~\rm{cm^{-3}}$. The scattering rates for distorted $\rm Cs_3Sb$ have been taken from  Ref. \citenum{Aryal2025} for comparison.}
\label{fig:Fig2a}
\end{figure}

Figure \ref{fig:Fig2a} illustrates the scattering rates for electrons and holes at room temperature ($T=300$ K) and a doping concentration of $10^{16}~\rm{cm^{-3}}$ for ordered $\rm CsK_2Sb$, distorted $\rm Cs_3Sb$, and distorted $\rm K_3Sb$. Our calculations predict that the electron scattering rate in these materials, near the conduction band minimum (CBM), is approximately $10^{14}~\rm{s^{-1}}$.  
Among them, $\rm Cs_3Sb$ and $\rm CsK_2Sb$ exhibit a slightly higher electron scattering rate near the CBM than $\rm K_3Sb$. In contrast, the hole scattering rate follows the opposite trend: $\rm K_3Sb$ has a slightly higher hole scattering rate than $\rm CsK_2Sb$ in the vicinity of valence band maximum (VBM), which is marginally higher than that of $\rm Cs_3Sb$.
The hole scattering rate near the VBM for these materials is approximately $10^{15}~\rm{s^{-1}}$, suggesting that the hole lifetime is shorter than the electron lifetime in the alkali-metal-antimonides considered in this work.  
Our calculations also indicate that POP scattering contributes almost entirely to the total scattering rates of both electrons and holes, except at the immediate band edges, where IMP scattering also plays a significant role. The dominant contribution of POP scattering in alkali-metal antimonides can be attributed to the large electronegativity difference between alkali metal and antimony atoms in these materials.
\begin{figure}[ht!]
\centering
\includegraphics[width=12cm]{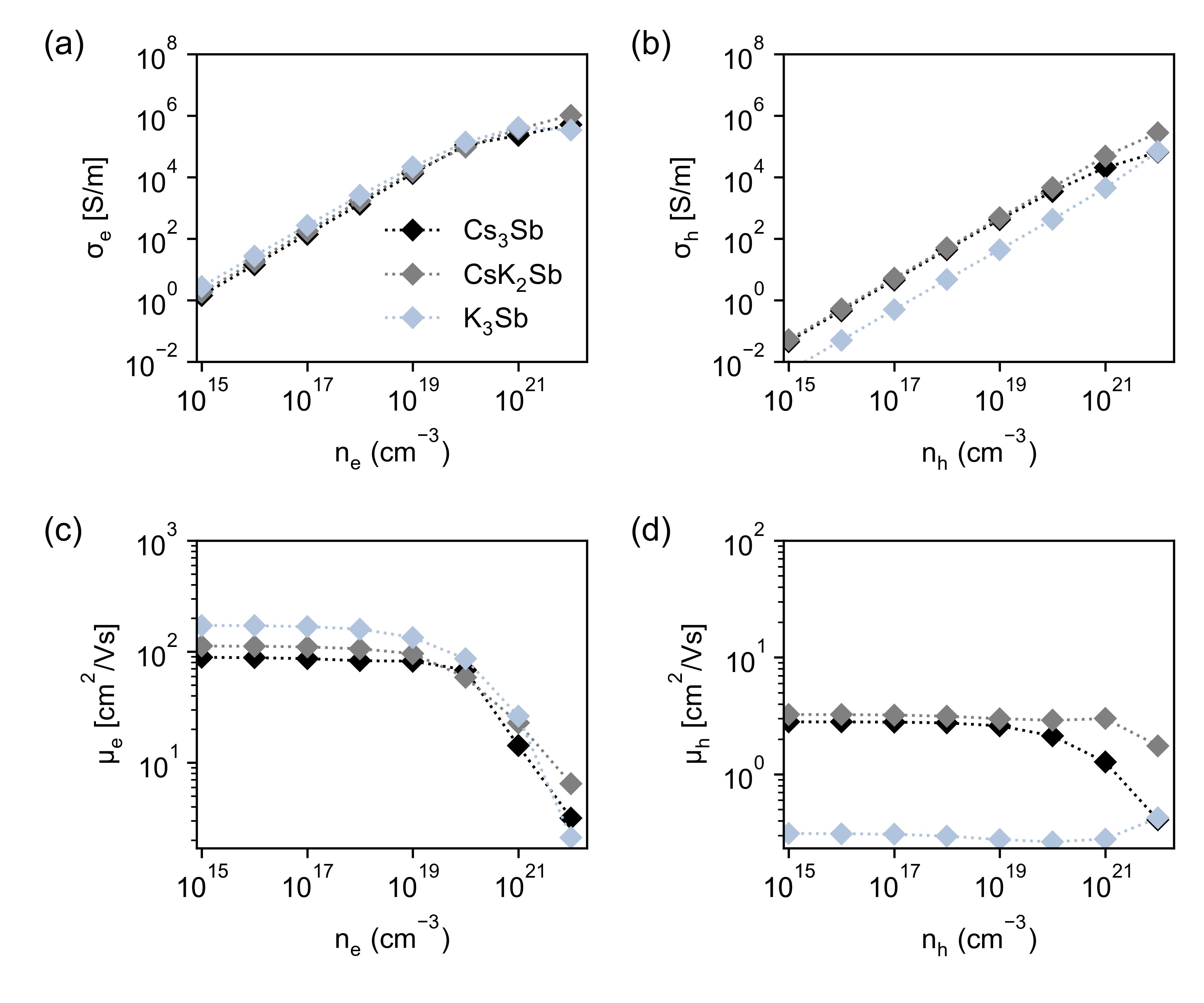}
\caption{Electrical conductivity and mobility of electrons and holes as a function of doping concentration for distorted $\rm Cs_3Sb$, ordered $\rm CsK_2Sb$, and distorted $\rm K_3Sb$ at $\rm T = 300$ K. (a) and (c) represent n-type (electron) doping, while (b) and (d) represent p-type (hole) doping.  The conductivity and mobility for distorted $\rm Cs_3Sb$ have been taken from  Ref. \citenum{Aryal2025} for comparison.}

\label{fig:Fig2}
\end{figure}

To gain deeper insight into carrier transport in these materials, we examine the conductivity and mobility of both electrons (n-type) and holes (p-type). Figure \ref{fig:Fig2} presents the electrical conductivity and mobility of electrons and holes as a function of doping concentration at $T = 300$ K.  
Both n-type (electron) and p-type (hole) conductivity increase with dopant concentration due to the greater number of carriers available for conduction in the valence and conduction bands. The mobility of electrons and holes remains nearly constant below a doping concentration of $10^{18}~\rm{cm^{-3}}$. However, above $10^{18}~\rm{cm^{-3}}$, both electron and hole mobility decrease with increasing carrier concentration. This decline is attributed to the rise in IMP scattering, in addition to POP scattering, as doping concentration increases (see Figure {\color{blue}{S2}}). 
\begin{figure}[ht!]
\centering
\includegraphics[width=12cm]{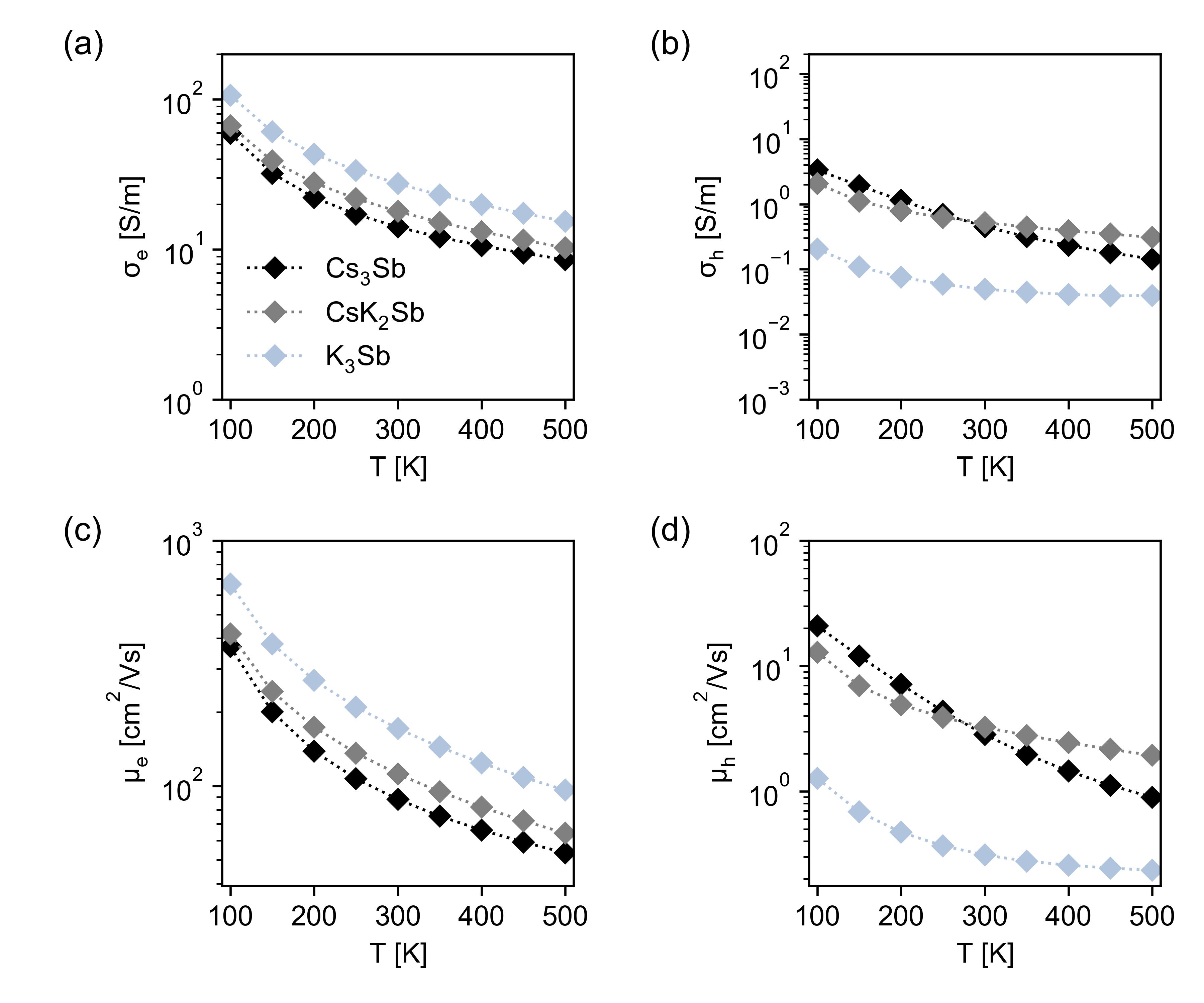}
\caption{Electrical conductivity and mobility of electrons and holes as a function of temperature for distorted $\rm Cs_3Sb$, ordered $\rm CsK_2Sb$, and distorted $\rm K_3Sb$ at a doping concentration of $10^{16}~\rm{cm^{-3}}$. (a) and (c) represent n-type (electron) doping, while (b) and (d) represent p-type (hole) doping.  The conductivity and mobility for distorted $\rm Cs_3Sb$ have been taken from  Ref. \citenum{Aryal2025} for comparison.
}
\label{fig:Fig3}
\end{figure}

Our calculations further reveal that $\rm K_3Sb$ exhibits higher electron conductivity and mobility compared to $\rm CsK_2Sb$ and $\rm Cs_3Sb$. This is likely due to its lower electron scattering rate, which results in a longer electron lifetime. In contrast, $\rm K_3Sb$ has lower hole conductivity and mobility than $\rm CsK_2Sb$ and $\rm Cs_3Sb$, attributed to its higher hole scattering rates, leading to a significantly shorter hole lifetime.  At $T = 300$ K and an electron and hole doping concentration of $10^{16}~\rm{cm^{-3}}$, the n-type conductivity for $\rm K_3Sb$ is 27.43 S/m, compared to 17.92 S/m for $\rm CsK_2Sb$ and 14.12 S/m for $\rm Cs_3Sb$. Similarly, the n-type mobility for $\rm K_3Sb$ is 171.22 $\rm{cm^2/Vs}$, compared to 111.86 $\rm{cm^2/Vs}$ for $\rm CsK_2Sb$ and 88.14 $\rm{cm^2/Vs}$ for $\rm Cs_3Sb$.  
For p-type conductivity, $\rm K_3Sb$ exhibits a significantly lower value of 0.050 S/m, compared to 0.52 S/m for $\rm CsK_2Sb$ and 0.45 S/m for $\rm Cs_3Sb$. Similarly, the p-type mobility for $\rm K_3Sb$ is 0.31 $\rm{cm^2/Vs}$, whereas it is 3.24 $\rm{cm^2/Vs}$ for $\rm CsK_2Sb$ and 2.81 $\rm{cm^2/Vs}$ for $\rm Cs_3Sb$.  
The lower conductivity and mobility of holes compared to electrons can be attributed to: (i) the larger effective mass of holes relative to electrons, and  (ii) the higher scattering rates for holes than for electrons.

Figure \ref{fig:Fig3} illustrates the temperature dependence of electrical conductivity and mobility at a doping concentration of $10^{16}~\rm{cm^{-3}}$. Both conductivity and mobility decrease with increasing temperature, primarily due to enhanced electron-phonon scattering as temperature rises.  
Figure \ref{fig:Fig3} also shows that the n-type conductivity and mobility of $\rm CsK_2Sb$ fall between those of distorted $\rm Cs_3Sb$ and distorted $\rm K_3Sb$ across the entire temperature range considered in this study. The p-type conductivity and mobility of $\rm CsK_2Sb$ exhibit higher values than those of distorted $\rm Cs_3Sb$ and distorted $\rm K_3Sb$ above room temperature. However, below room temperature, the p-type conductivity and mobility of $\rm CsK_2Sb$ lie between those of the other two materials.
\subsection*{3.4 Native Defects in $\rm CsK_2Sb$}
Point defects, such as vacancies, interstitials, and antisites, inevitably form during the crystal growth process at any finite temperature under equilibrium conditions. These defects can significantly influence the photoemission properties of a material by modifying its optical and transport properties, as well as the escape rate of electrons from its surface. Therefore, we investigate the stability, electronic, and optical properties of $\rm CsK_2Sb$ in the presence of native point defects. For this, a large $2\times2\times2$ supercell of $\rm CsK_2Sb$ with a lattice constant of $\sim 17.12$~\AA, consisting of 32 Cs, 64 K, and 32 Sb atoms, was constructed from its conventional unit cell shown in Figure \ref{fig:Fig1a}(b). The rationale for using such a large supercell stems from the need to maintain a low defect concentration in our simulations. Additionally, it minimizes defect-defect interactions between a defect and its periodic images in  calculations. 

Various types of defects, including mono- and di-vacancies of Cs, K, and Sb, as well as their antisites and interstitials, were introduced into the supercell by removing or adding atoms. In this work, a mono-vacancy is denoted as $\rm V_X$ (where $\rm X = Cs, K,$ or $\rm Sb$), indicating a missing atom at the $X$ site. Similarly, a di-vacancy formed by the removal of two identical atoms is represented as $\rm V_{2X}$ (where $\rm X = Cs, K,$ or $\rm Sb$), signifying the simultaneous absence of two atoms at the $X$ site. If the di-vacancy is formed from two dissimilar atoms, it is denoted as $\rm V_{XY}$ (where $\rm X, Y = Cs, K,$ or $\rm Sb$), indicating the simultaneous absence of atoms at both the $X$ and $Y$ sites.
The antisite defect, in which an Sb atom occupies a Cs site, is denoted as $\rm Sb_{Cs}$. Other antisite defects are labeled in the same manner. The defect supercells were then relaxed at the PBE+D3 level to obtain the equilibrium lattice constants and atomic positions.  
Our calculations show that the lattice constants of the defect supercells change by an insignificant amount (${\pm}0.7\%$) upon relaxation compared to the pristine lattice constant.
\subsubsection*{Defect formation energy}
The energy cost of forming a defect in a neutral charge state in a material can be quantified by its formation energy, $\rm E_F$, which is defined as:
\begin{equation}
\rm E_F(\mu_{Cs}, \mu_{K}, \mu_{Sb}) = E_D - E_P + n_{Cs}\mu_{Cs}  + n_{K}\mu_{K} + n_{Sb}\mu_{Sb}
\label{E1}
\end{equation}
Here, $\rm E_D$ and $\rm E_P$ represent the total energy of the supercell with and without a defect, respectively. The term \( n_i \) (where $\rm i = \text{Cs}, \text{K}, and \hspace{0.1cm}\text{Sb} $) denotes the number of atoms either added or removed to introduce a defect, while \( \mu_i \) corresponds to the chemical potential of species i (where $\rm i = \text{Cs}, \text{K}, and \hspace{0.1cm}\text{Sb} $). A positive \( n_i \) (\( n_i > 0 \)) indicates the removal of an atom from the system, whereas a negative \( n_i \) (\( n_i < 0 \)) signifies the addition of an atom.
Under thermodynamic equilibrium conditions, the chemical potentials of Cs (\( \mu_{\rm Cs} \)), K (\( \mu_{\rm K} \)), and Sb (\( \mu_{\rm Sb} \)) must satisfy the following condition:
\begin{equation}
\rm  \mu_{CsK_2Sb} = \mu_{Cs} + 2\mu_{K} + \mu_{Sb}
\label{E2}
\end{equation}
where, \( \mu_{\rm CsK_2Sb} \) represents the energy per formula unit of $\rm CsK_2Sb$, denoted as \( E_{\rm CsK_2Sb}^{\rm bulk} \). The formation energy in Equation \(\ref{E1}\) is a function of three variables: \( \mu_{\rm Cs} \), \( \mu_{\rm K} \), and \( \mu_{\rm Sb} \). However, since these chemical potentials are constrained by Equation \(\ref{E2}\), the formation energy can be expressed in terms of any two of these variables. That is, $\rm E_F(\mu_{\rm Cs}, \mu_{\rm K}, \mu_{\rm Sb}) \sim E_F(\mu_{\rm Cs}, \mu_{\rm K})$. With this substitution, Equation \(\ref{E1}\) simplifies to:
\begin{equation}
\rm  E_F(\mu_{Cs}, \mu_{K}) = E_D - E_P + n_{Cs}\mu_{Cs}  + n_{K}\mu_{K} + n_{Sb}(\mu_{CsK_2Sb} - \mu_{Cs} - 2\mu_{K})
\label{E3a}
\end{equation}
We study the formation energy as a function of $\rm \mu_K$ for different values of $\rm \mu_{Cs}$.
The elemental chemical potential of the species must satisfy the following inequality:
%
%
\begin{align}
\rm E_{CsK_2Sb}^{bulk} - 2\rm E_{K}^{atom} -E_{Sb}^{atom} < \rm \mu_{Cs} < E_{Cs}^{atom}
\label{A1}
\end{align}
\begin{align}
\rm (E_{CsK_2Sb}^{bulk} - E_{Cs}^{atom} -E_{Sb}^{atom})/2 < \rm \mu_{K} < E_{K}^{atom}
\label{A2}
\end{align}
where, \( E_{\rm Cs}^{\rm atom} \), \( E_{\rm K}^{\rm atom} \), and \( E_{\rm Sb}^{\rm atom} \) are the cohesive energies per atom of Cs, K, and Sb atom in their bulk crystals, respectively. These energies were obtained using DFT in this study. The right-hand side of the inequality in Equation \(\ref{A1}\) corresponds to Cs-rich conditions, while the left-hand side represents Cs-poor conditions. Similarly, the right-hand side of the inequality in Equation \(\ref{A2}\) denotes K-rich conditions, whereas the left-hand side corresponds to K-poor conditions.  
Our calculations yield:
\begin{align}
\rm -2.73\hspace{0.1cm}eV < \rm \mu_{Cs} < -0.93\hspace{0.1cm}eV\nonumber \\
\rm -2.05\hspace{0.1cm}eV < \rm \mu_{K} < -1.15\hspace{0.1cm}eV
\label{A3}
\end{align}
The thermodynamic stability of defects in $\rm CsK_2Sb$ was analyzed within the range of chemical potentials, $\mu_{\rm Cs}$ and $\mu_{\rm K}$, as defined by the inequality in Equation \ref{A1}.

\begin{figure}[ht!]
\centering
\includegraphics[width=16cm]{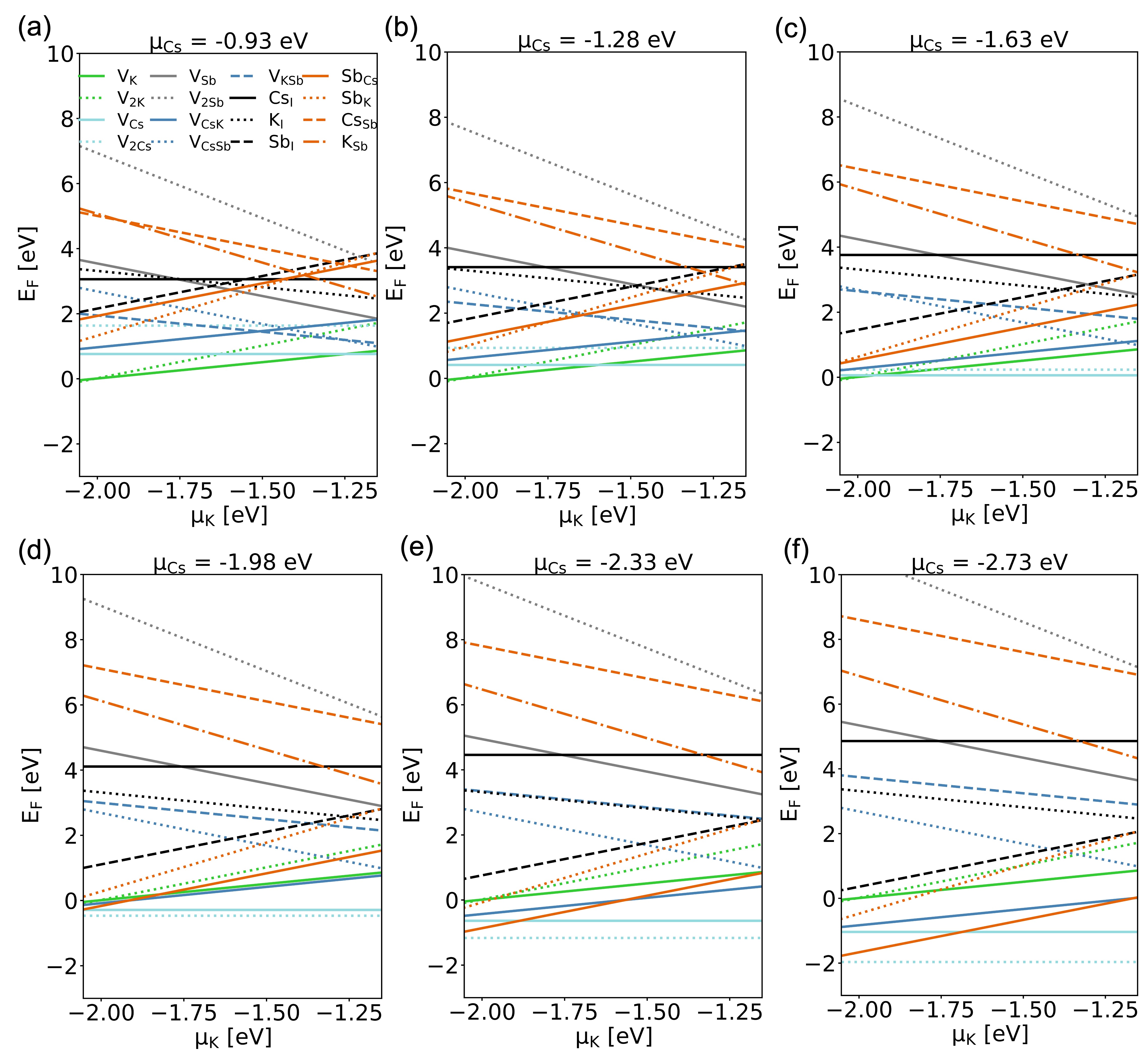}
\caption{(a–c) Defect formation energy as a function of $\mu_{\rm K}$ for different values of $\mu_{\rm Cs}$.}
\label{fig:Fig4}
\end{figure}

Figure \(\ref{fig:Fig4}\) presents the formation energy of native point defects as a function of $\mu_{\rm K}$ for six different values of $\mu_{\rm Cs}$, within the inequality defined in Equation \(\ref{A1}\).  
In the plot, $\mu_{\rm K} = -1.15$ eV represents K-rich conditions, whereas $\mu_{\rm K} = -2.05$ eV denotes K-poor conditions. Similarly, $\mu_{\rm Cs} = -0.93$ eV and $\mu_{\rm Cs} = -2.73$ eV correspond to Cs-rich and Cs-poor conditions, respectively.  
We first analyze the defect formation energy within the range -1.63 $<$ $\mu_{\rm Cs}$ $<$ -0.93, where the defect formation energy remains mostly positive over a wide range of $\mu_{\rm K}$. Within this range, Cs and K related vacancies and di-vacancies are energetically more favorable than other point defects across a broad range of $\mu_{\rm K}$.

Under Cs-rich and K-rich conditions, the formation of Cs vacancy,  $\rm V_{Cs}$, is energetically favorable. $\rm V_{K}$ remains stable below $\mu_{\rm K} \leq -1.25$ eV under Cs-rich conditions. Approximately below $\mu_{\rm K} \leq -2.01$ eV, i.e., in the vicinity of K-poor conditions, $\rm V_{2K}$ has a lower formation energy than $\rm V_{K}$, despite its formation energy being negative. The negative formation energy suggests the spontaneous formation of $\rm V_{2K}$ in the material that might lead to crystal instability if such defects are formed in higher concentration.  As $\mu_{\rm Cs}$ deviates from Cs-rich conditions, i.e., shifting to -1.28 eV and -1.63 eV, $\rm V_{Cs}$ maintains the lowest formation energy over a wider range of $\mu_{\rm K}$ compared to $\rm V_{K}$. When $\mu_{\rm Cs} = -1.28$ eV, $\rm V_{Cs}$ is found to be energetically favorable within the range $-1.60$ eV $\leq \mu_{\rm K} \leq -1.15$ eV, whereas $\rm V_{K}$ is more stable in the range $-2.05$ eV $\leq \mu_{\rm K} \leq -1.60$ eV. Similarly, for $\mu_{\rm Cs} = -1.63$ eV, $\rm V_{Cs}$ remains more stable within $-1.95$ eV $\leq \mu_{\rm K} \leq -0.93$ eV, while below this range, $\rm V_{K}$ and $\rm V_{2K}$ become energetically favorable.  
Furthermore, our calculations show that $\rm V_{Cs}$ and $\rm V_{2Cs}$ exhibit negative formation energies within the range $-2.73 < \mu_{\rm Cs} < -1.98$, suggesting that these vacancies are more likely to form than pristine $\rm CsK_2Sb$ during the experimental growth process under these conditions. The spontaneous formation of these defects can result in a non-stoichiometric composition, altering the overall properties of the material, and causing crystal instability if higher concentration of these defects are introduced in the material. However, within this region, other defects exhibit positive formation energies, depending on the chemical potential of $\mu_{\rm Cs}$. When \( \mu_{\rm Cs} = -1.98 \) eV, the Cs-K di-vacancy, $\rm V_{CsK}$, exhibits a lower positive formation energy than $\rm V_{K}$ and $\rm V_{2K}$ over a wide range of $\mu_K$. As \( \mu_{\rm Cs} = -2.33 \) eV, $\rm V_{CsK}$ and the $\text{Sb}_{\rm Cs}$ antisite defect have small but positive formation energy in the vicinity of K-rich conditions. Below $\rm \mu_K$ = -1.57 eV, the formation energy for these defects become negative. When \( \mu_{\rm Cs} = -2.73 \) eV, both $\rm V_{CsK}$ and the $\text{Sb}_{\rm Cs}$ antisite defect exhibit negative formation energies irrespective of the values of $\rm \mu_K$. This is much expected as the lack of Cs leads to formation of vacancies or those sites being populated by other species. We should mention that this is an extreme situation, where $\rm CsK_2Sb$ is competing with metallic Cs for Cs atoms, and not expected under laboratory conditions. Other defects, including Sb vacancies, Sb di-vacancies, interstitials of Cs, K, and Sb, as well as Cs and K antisite defects, exhibit relatively higher formation energies compared to these defects, regardless of \( \mu_{\rm Cs} \) and \( \mu_{\rm K} \).

\subsubsection*{Stability of vacancy related clusters}
Since the vacancy defects in $\rm CsK_2Sb$ are relatively favorable than other defects, to evaluate the stability of a vacancy cluster, we examine its binding energy. In analogy with the definition of molecular atomization energies, the binding energy (\(\rm E_b\)) of a defect cluster was defined from the formation energies of the cluster and the formation energy of the individual defects:
\begin{equation}
\rm E_b = E_F^{cluster} - \sum E_F^{isolated}
\end{equation}
where \(\rm E_F^{\rm complex} \) represents the formation energy of the defect cluster, \( E_F^{\rm isolated} \) denotes the formation energy of isolated defects, and the summation accounts for all defects contributing to the cluster.  
A cluster will be stable, relative to the individual defects, if \( E_b < 0 \), and unstable if \( E_b > 0 \). In the second case, it will tend to thermally dissociate.  
Table \ref{tab:table3} presents the calculated binding energy for the vacancy clusters in $\rm CsK_2Sb$.
\begin{table}[ht!]
\caption{\label{tab:table3} Computed binding energy of vacancy cluster in $\rm CsK_2Sb$.}
\centering
\begin{tabular}{c c c c c c c} 
\hline\hline
& $\rm V_{CsCs}$ & $\rm V_{CsK}$ & $\rm V_{KK}$ &  $\rm V_{SbSb}$ & $\rm V_{KSb}$ & $\rm V_{CsSb}$ \\
\hline
$\rm E_b$ (eV) & 0.11 & 0.20 & -0.0005 & -0.15 & -1.60 & -1.62  \\
\hline\hline
\end{tabular}
\end{table}

Our calculations indicate that the di-vacancy clusters, Cs-Cs and Cs-K, are unstable in spite of their relatively low formation energies compared to other native defects. Similarly, while the \( \rm V_{2K} \) cluster also has a relatively low formation energy, the weak attractive interactions between K vacancies suggest its instability at higher temperature.  
The strongest binding is observed for di-vacancies formed by the removal of a cation and an anion in $\rm CsK_2Sb$, specifically \( \rm V_{KSb} \) and \( \rm V_{CsSb} \). However, these complexes are difficult to form due to their relatively high formation energies. Nevertheless, if formed, they are expected to remain stable.  
Likewise, the \( \rm V_{2Sb} \) cluster is found to be stable due to its negative binding energy. However, its formation energy is significantly high in \( \rm CsK_2Sb \), making its occurrence less likely.
\subsubsection*{Electronic density of states}
Since vacancy defects are the dominant defects in $\rm CsK_2Sb$, we investigate their effect on  electronic properties. Our calculations indicate that the presence of vacancies at low concentrations in $\rm CsK_2Sb$ does not significantly alter the overall shape of the valence and conduction bands (see Figure {\color{blue}{S3}}). However, these vacancies introduce electronic states within the bandgap of $\rm CsK_2Sb$. Unlike Cs- and K-related vacancies, which induce unoccupied states above the Fermi level in the vicinity of the VBM, the Sb-related mono- and di-vacancies introduce states below the Fermi level in the proximity of the CBM and within the middle of the bandgap.
Defect states located closer to the band edges act as shallow traps, whereas those further from the band edges serve as deep traps.  In general, deep traps are detrimental as they contribute to charge trapping and non-radiative recombination, thereby impacting the optical properties, charge transport, and carrier ejection from the surface of $\rm CsK_2Sb$. Consequently, these effects influence the photoemission process in $\rm CsK_2Sb$ photocathodes.

Figure {\color{blue}{S3}} further illustrates that the presence of defects shifts the energy bands and the Fermi level relative to those of pristine $\rm CsK_2Sb$. For instance, in defective $\rm CsK_2Sb$ with a single Cs vacancy, both the VBM and the Fermi level shift leftward. 
As the concentration of Cs vacancies increases, the Fermi level moves further into the valence band, with the VBM also shifting left, as shown in Figure {\color{blue}{S3}}. This represents ``hole-doped'' (p-type conduction) in the sense that the Fermi level moves inside the valence band. A similar trend is observed for K-related mono- and di-vacancies, and $\rm V_{CsK}$ di-vacancy.
For the Sb-related vacancy and di-vacancies, the Fermi level moves inside the conduction band, signifying the ``electron-doped'' (n-type conduction) in non-stoichimetric $\rm CsK_2Sb$.

%
\subsubsection*{Optical properties}
The first step in the Spicer model of photoemission is determined by the response to light by the material, specifically, the absorption efficiency. The presence of native point defects in $\rm CsK_2Sb$ is expected to influence its photoemission, as these defects introduce traps within the bandgap that can alter its optical properties. To examine the impact of defects on the optical response of $\rm CsK_2Sb$, we computed its optical absorption coefficient as a function of photon energy for defective $\rm CsK_2Sb$ and compared the results with those of the pristine material. In this work, we focus on vacancy defects, as these are the expected primary defects (see subsection 3.4.).
%
\begin{figure}[ht!]
\centering
\includegraphics[width=18cm]{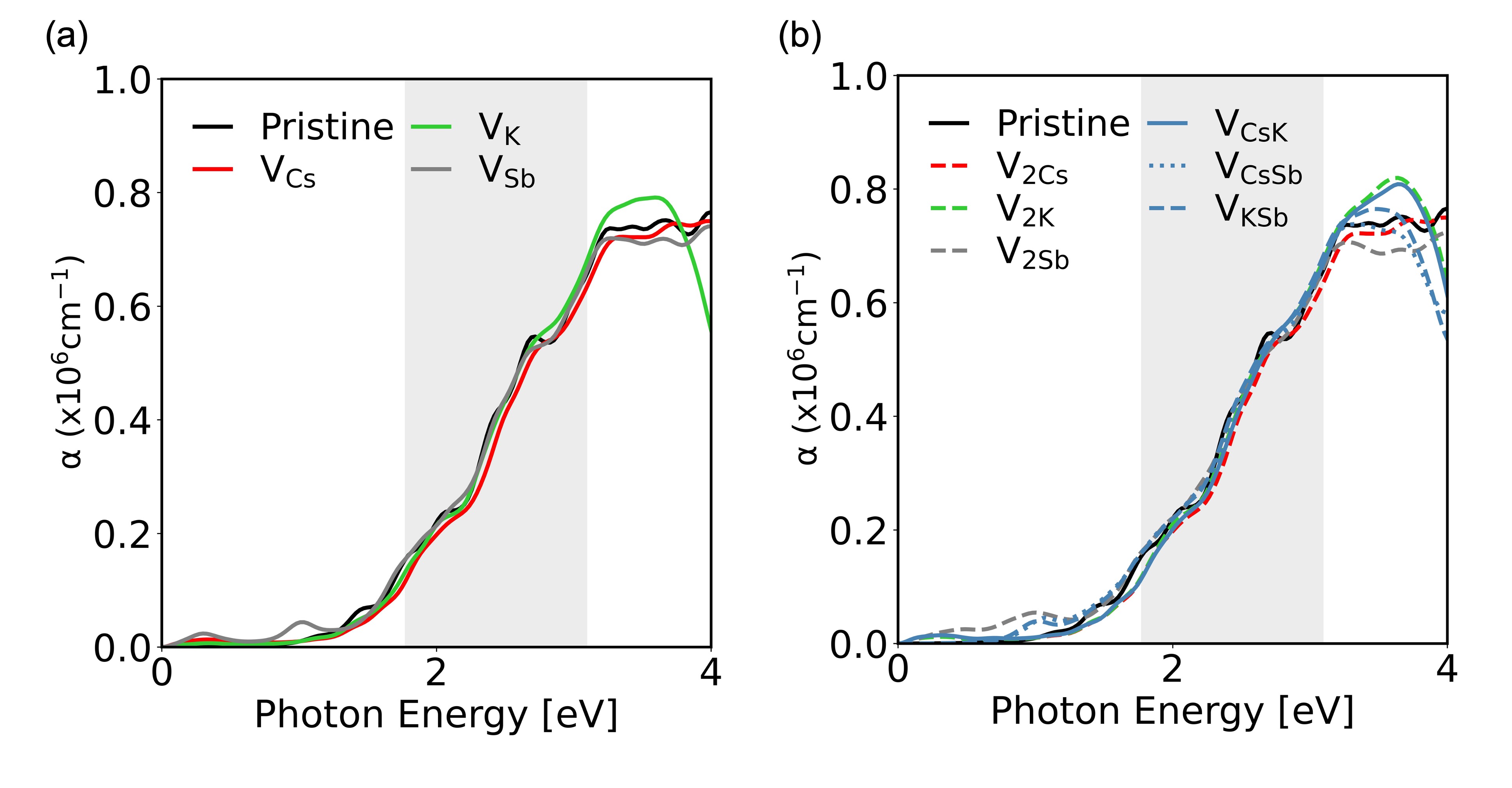}
\caption{Optical absorption as a function of photon energy for defective $\rm CsK_2Sb$ (a) mono-vacancy, and (b) di-vacancies formed by removal of Cs, K, and Sb atoms. The shaded region in the plot represents the visible spectrum.}
\label{fig:Fig5}
\end{figure}

The computed optical absorption coefficient (Figure \ref{fig:Fig5}) for pristine $\rm CsK_2Sb$ ranges from $\rm 1.52 \times 10^{5} \leq \alpha (cm^{-1}) \leq \rm 6.43 \times 10^{5}$, suggesting strong absorption in the visible range. This is further supported by the observed high quantum yield of (8–12)\% for $\rm CsK_2Sb$ at 527–532 nm. \cite{WRB2014, LJID2011, DSK1995}  The calculated optical absorption for pristine $\rm CsK_2Sb$ agrees well with the experimental absorption coefficient of ($\sim 5 \times 10^5$ $\rm cm^{-1}$ at 3 eV),\cite{S581, G9845} obtained for alkali-metal-antimonide photocathodes. The presence of vacancy defects in $\rm CsK_2Sb$ induces low-energy peaks in the absorption spectrum below the bandgap, as seen in Figure \ref{fig:Fig5}(a-b). The maximum change in the absorption coefficient at a photon energy of 2.5 eV, relative to the pristine value, is approximately $\pm$4\%. Additionally, our calculations show that the change in the absorption coefficient becomes significant beyond a photon energy of 3.2 eV. At 3.6 eV, the maximum change in the absorption coefficient reaches approximately $\pm$10\%. Such significant variations in absorption influence the quality of the electron beam generated by the $\rm CsK_2Sb$ photocathode, thereby affecting its overall performance.
\subsection*{3.5 Surface Properties of $\rm CsK_2Sb$}
According to Spicer’s three-step model of photoemission, the final stage—electron escape—is strongly influenced by the surface properties of a material. Electron emission from the surface is directly affected by several key factors, including surface stability, work-function/ionization potential, surface roughness, and presence of adsorbates. Consequently, a detailed understanding of surface properties is essential for optimizing photoemission performance. This study focuses on evaluating relative surface formation energies, electronic DOS, and work-function for low Miller index crystallographic facets, specifically (100), (110), and (111). The objective is to identify the most stable surface orientation for optimal electron emission in $\rm CsK_2Sb$.
\subsubsection*{Surface stability}
 The stability of the slabs was investigated by calculating the surface energy defined as:
\begin{equation}
\rm \gamma(\mu_{Cs}, \mu_{K}, \mu_{Sb}) = \frac{1}{2A}[E_{slab} - n_{Cs}\;\mu_{Cs} - n_{K}\;\mu_{K} - n_{Sb}\;\mu_{Sb}]
\label{ES1}
\end{equation}
where \( A \) and \( E_{\rm slab} \) are the surface area and total energy of the slab, respectively. The terms \( n_{\rm Cs} \), \( n_{\rm K} \), and \( n_{\rm Sb} \) denote the number of Cs, K, and Sb atoms in the slab, while \( \mu_{\rm Cs} \), \( \mu_{\rm K} \), and \( \mu_{\rm Sb} \) correspond to their respective chemical potentials. Similar to defect formation energy, the surface energy of \( \rm CsK_2Sb \) is a function of three variables: \( \mu_{\rm Cs} \), \( \mu_{\rm K} \), and \( \mu_{\rm Sb} \). However, one of these variables (\( \mu_{\rm Sb} \)) can be eliminated from Equation \ref{ES1} using the thermodynamic equilibrium condition given by Equation \ref{E2}.
In terms of the chemical potential of the Cs and K, the surface energy in Equation \ref{ES1} can be expressed as
\begin{equation}
 \rm  \gamma(\mu_{Cs}, \mu_{K}) = \frac{1}{2A}[E_{slab} - \mu_{Cs}(n_{Cs}-n_{Sb}) - \mu_{K}(n_{K} - 2n_{Sb}) - n_{Sb}\mu_{CsK_2Sb}]
\label{ES3}
\end{equation}
Thus, we reduced a function of three variables to two. The elemental chemical potentials, \( \mu_{\rm Cs} \) and \( \mu_{\rm K} \), in Equation \ref{ES3} are calculated from their respective bulk phases.  
In this work, we computed the surface energy of different \( \rm CsK_2Sb \) facets within the limits of \( \mu_{\rm Cs} \) and \( \mu_{\rm K} \) as defined in Equations \ref{A1}, \ref{A2}, and \ref{A3}.
\begin{figure}[ht!]
\centering
\includegraphics[width=12cm]{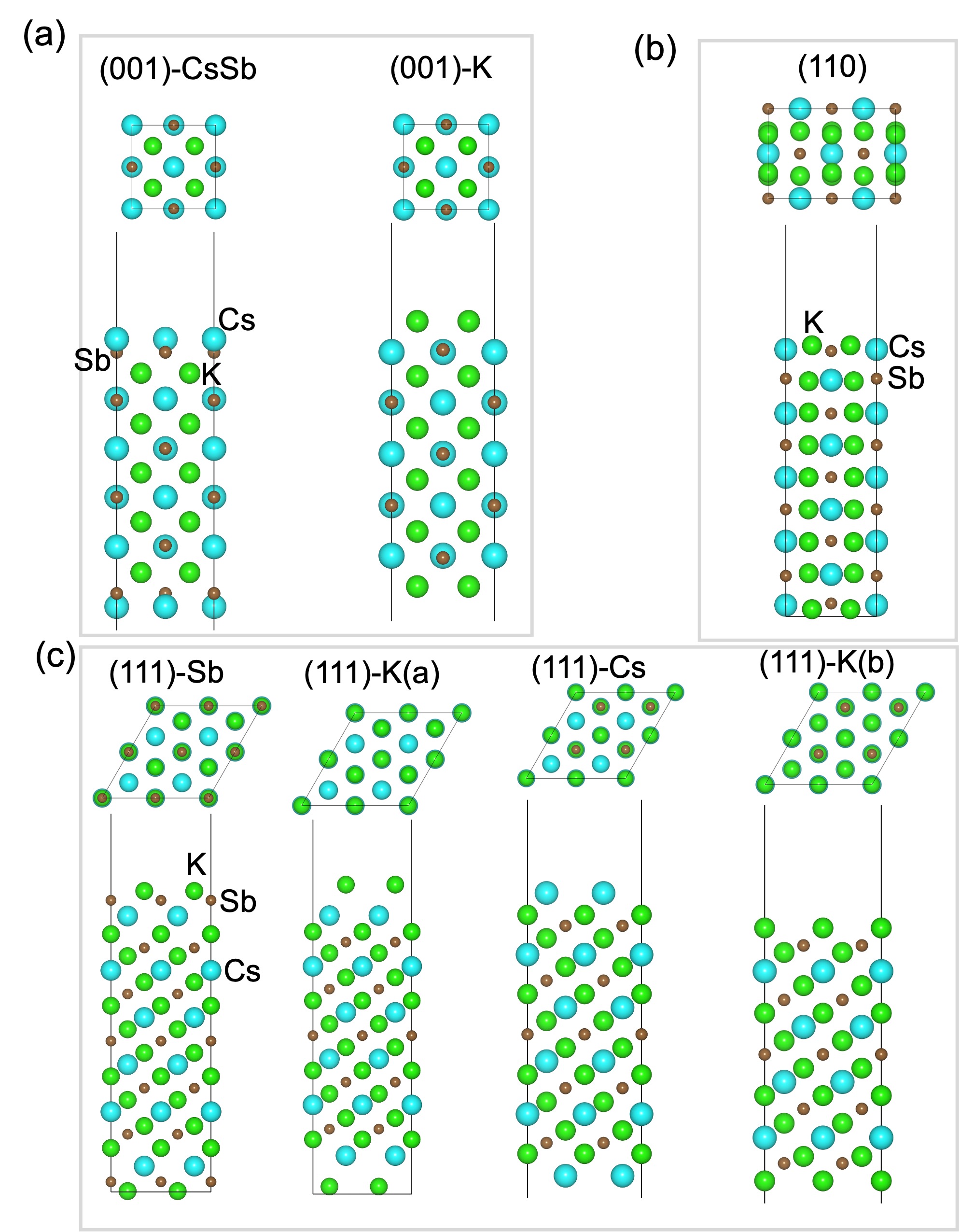}
\caption{(a–c) Relaxed unit cell structures (vertical and top views) of the (001), (110), and (111) surfaces with different terminations considered in this study. Cs, K, and Sb atoms are represented by cyan, green, and brown spheres, respectively.}
\label{fig:Fig6}
\end{figure}

\begin{figure}[ht!]
\centering
\includegraphics[width=16cm]{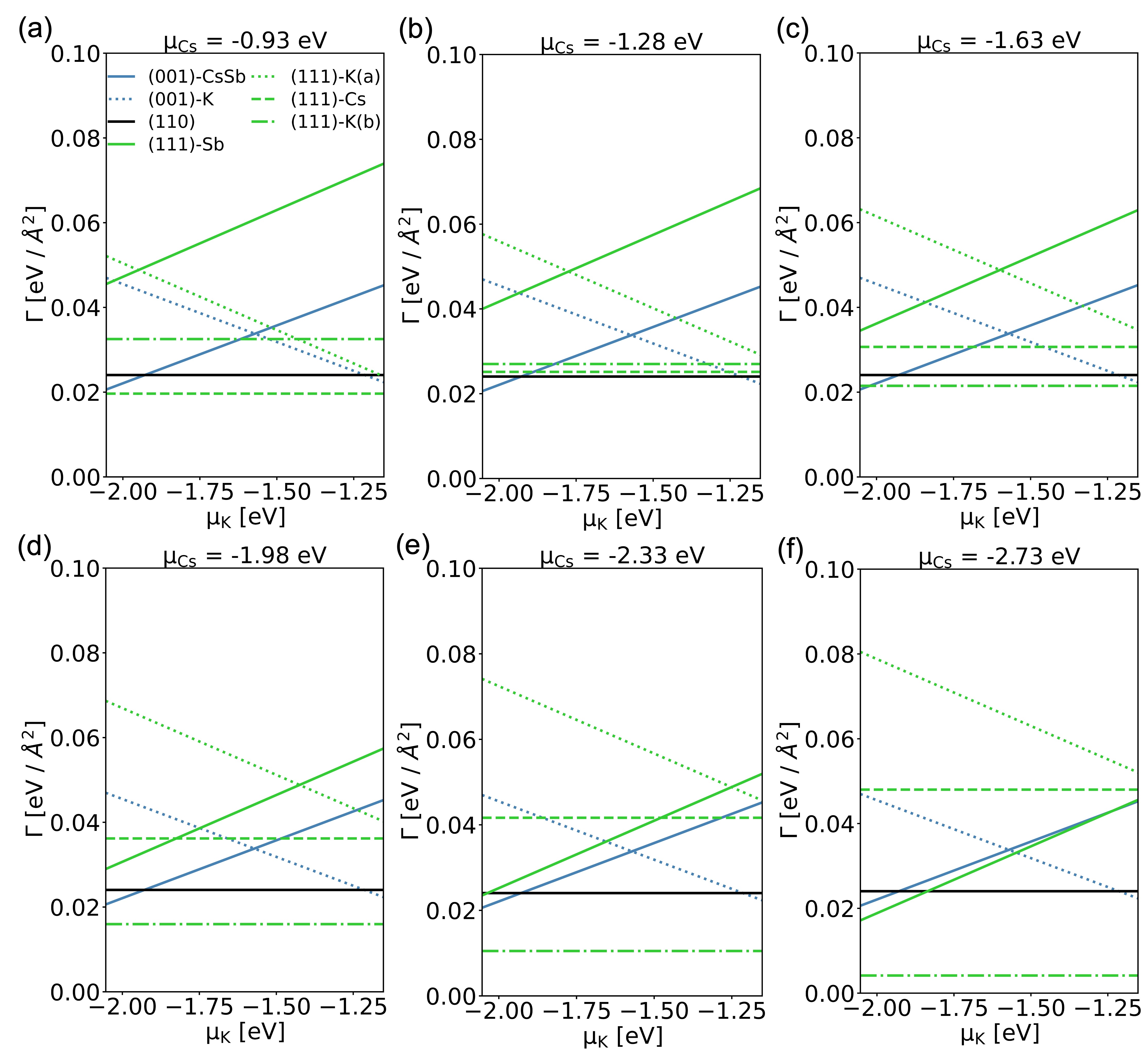}
\caption{(a-c) Surface energy as a function of $\rm{\mu_{K}}$ for different $\rm{\mu_{Cs}}$.}
\label{fig:Fig7}
\end{figure}

Figure \ref{fig:Fig6} illustrates the relaxed slabs, each with identical terminations on both sides.
Notice that, upon optimization, the (001)-CsSb slab distorts with the Cs atoms relaxing outwards creating a Cs/Sb bilayer and inducing an electric dipole polarization of the surface. This reorganization mostly affects the top layer, with minimal distortion on the second layer. For simplicity, we will continue to refer to this as the (001)-CsSb surface in Figure \ref{fig:Fig6}(a).  Similar relaxation and induced surface polarization was observed when cleaving and optimizing the Sb termination of the (111) plane. Upon optimization the K sub-surface layer spontaneously relaxes outwards becoming the top layer and the Sb atoms the sub-surface layer, as shown in Figure \ref{fig:Fig6}(c). For consistency this surface will still be labeled as (111)-Sb. Two K-terminated surfaces, denoted as (111)-K(a) and (111)-K(b), were considered in this study and are shown in Figure \ref{fig:Fig6}(c). In the (111)-K(a) slab, the surface K layer tends to separate from the layers beneath it, as depicted in Figure \ref{fig:Fig6}(c). However, this surface is less stable due to its higher surface energy (see Figure \ref{fig:Fig7}).

Figure \ref{fig:Fig7} presents the surface energy of $\rm CsK_2Sb$ surface slabs as a function of $\mu_{\rm K}$ for different values of $\mu_{\rm Cs}$. As shown in Figure \ref{fig:Fig7}, the stability of the $\rm CsK_2Sb$ surface slabs is highly dependent on $\mu_{\rm Cs}$ and $\mu_{\rm K}$.  
In Figure \ref{fig:Fig7}, $\mu_{\rm K} = -1.15 \, \text{eV}$ represents K-rich conditions, while $\mu_{\rm K} = -2.05 \, \text{eV}$ corresponds to K-poor conditions. Similarly, $\mu_{\rm Cs} = -0.93 \, \text{eV}$ and $\mu_{\rm Cs} = -2.73 \, \text{eV}$ denote Cs-rich and Cs-poor conditions, respectively. These values define the chemical potential limits for K and Cs under different environmental conditions.  
Our calculations reveal that under Cs-rich conditions (Fig \ref{fig:Fig7}(a)), irrespective of $\mu_{\rm K}$, the (111)-Cs surface is the most stable due to its lowest surface energy of approximately 0.0196 eV/{\AA}$^2$. This suggests that the (111)-Cs facet is the most exposed surface during crystal growth under Cs-rich conditions.  
By $\mu_{\rm Cs} = -1.28$ eV, as the presence of Cs decreases, the (110) surface is the most stable. This changes in the very K rich regimen, $\mu_{\rm K} \geq -1.21$ eV where the (001)-K surface is energetically the most favorable and in the K very poor regimen, $\mu_{\rm K} \leq -1.93$ eV, where the (001)-CsSb surface appears as the most stable. As $\mu_{Cs}$ decreases, mimicking a smaller presence of Cs, the (111)-K(b) surface appears as the most stable. 

To gain a better intuition of the relative stability of various $\mathrm{CsK_2Sb}$ surface terminations, a false-color plot (Figure {\color{blue}{S4}}) has been generated to map the thermodynamically favorable regions as a function of the chemical potentials $\mu_{\mathrm{Cs}}$ and $\mu_{\mathrm{K}}$. Each region in the plot corresponds to the surface termination with the lowest surface energy under specific chemical potential conditions. Figure {\color{blue}{S4}} reveal that the (111)-K(b) surface termination is stable across a broad range of $\mu_{\mathrm{Cs}}$ and $\mu_{\mathrm{K}}$ values. This finding is in good agreement with prior work  (Ref. [\citenum{Schier_2022}]), which employed the SCAN functional. This is indicative of the fact that the surface physics and crystal growth in this system being very sensitive to the experimental conditions, which should allow one to control the growth by varying the relative pressures of K and Cs in the chamber.

\subsubsection*{Density of states}
Figure {\color{blue}{S4}} presents the atom-decomposed DOS, highlighting the contributions of Cs, K, and Sb atoms. Our calculations reveal that the (001) and (111) slabs exhibit finite DOS at the Fermi energy, indicating their metallic behavior. Since the bulk crystal is semiconductor with a bandgap of 1.13 eV, the metallic nature of these slabs can be attributed to the formation of dangling bonds due to the under coordinated atoms on the surface upon cleaving. These dangling bonds introduce states within the bulk bandgap, resulting in a finite DOS at the Fermi level. 
Our analysis of DOS (Figure {\color{blue}{S5}}) suggests a strong mixing of the orbitals of Cs, K, and Sb that leads to such surface states and hence the metallic behavior.
In contrast, the (110) slab is found to be semiconductor, with an energy gap of 0.60 eV. The valence band of the (110) slab is primarily contributed by Sb atoms, with a small contribution from Cs and K atoms. Conversely, the conduction band is dominated by Cs and K atoms, with a minor contribution from Sb atoms. 

It is worth noting that the stability and electronic density of states of the $\mathrm{CsK_2Sb}$ (111) facets were also investigated in Ref. [\citenum{Schier_2022}] using the SCAN functional. While the computed surface energies in both studies fall within a comparable range, our results predict all four (111) surface terminations to be metallic. In contrast, Ref. [\citenum{Schier_2022}] identifies the (111)-K(b) and (111)-Sb terminations as semiconducting, whereas the (111)-K(a) and (111)-Cs terminations are found to be metallic. The calculated work function and/or ionization potential obtained in this work show reasonable agreement with those reported in Refs. [\citenum{Schier_2022}, \citenum{Wang_Zhang_2022}].
\subsubsection*{Work-function / Ionization potential}
The work-function of the material is the key parameter determining the efficiency of the third step in Spicer’s three-step model, that is, the electron escape from the surface. The work-function, WF, represent the minimum energy required for electron emission when the material is exposed to light. 
In metals, the work-function is referenced from the Fermi energy, as it is well-defined. Mathematically, it is expressed as:  
\begin{equation}
\rm \text{WF} = E_{\text{vac}} - E_{F}
\end{equation}
where \( E_{\text{vac}} \) is the vacuum energy level, and $\rm E_F$ is the Fermi energy.  

However, in semiconductors and insulators, where the Fermi energy is not well-defined, the ionization potential is usually computed, which is referenced to the valence band maximum (VBM). It is calculated as:  
\begin{equation}
\rm \text{IP} = E_{\text{vac}} - E_{\text{VBM}}
\end{equation}
where $\rm E_{\text{VBM}}$ represents the valence band maximum.  
A lower work-function or ionization potential enhances photoemission in materials by allowing electrons to escape more readily, which is crucial for applications requiring high quantum efficiency (QE), such as photodetectors, image intensifiers, and free-electron lasers. 
\begin{table}[ht!]
\caption{\label{tab:table2} Computed energy gap ($\rm E_g$), work-function (WF), and ionization potential (IP) for the slabs. The work-function is calculated for the metallic slabs, whereas IP and EA is calculated for the semiconducting.}
\centering
\begin{tabular}{c c c c c c c c} 
\hline\hline
&(001)-CsSb & (001)-K & (110) & (111)-Sb & (111)-K(a) & (111)-Cs & (111)-K(b) \\
\hline
$\rm E_g$ (eV) & - & - & 0.60 & - & - & - & - \\
WF (eV) & 2.78 & 2.54 & - & 3.29 & 2.21 & 2.14 &  2.56\\
IP (eV) & - & - & 2.64 & - & - & - & - \\
\hline\hline
\end{tabular}
\end{table}

Table \ref{tab:table2} presents the calculated energy gap, work-function, and ionization potential for the various surface terminations listed above. The lowest work-function, approximately 2.14 eV, is found for the Cs-terminated (111)-Cs surface. 
Similarly, the K-terminated (111)-K(a) and (111)-K(b) surfaces exhibit work-functions of 2.21 eV and 2.56 eV, while the (001)-CsSb surface has the work-function of 2.78 eV. The work-function calculated for the stable (111)-Cs and (111)-K(b) surfaces is significantly lower than that of widely used metallic photocathode materials such as Cu (WF = 4.65 eV), Mg (WF = 3.66 eV), and Pb (WF = 4.25 eV) \cite{M1997}. Furthermore, the work-function of $\rm CsK_2Sb$ is comparable to that of $\rm Cs_3Sb$ (2.01 eV) \cite{WYNB2018}, another promising photocathode material in the alkali-metal-antimonide family. The low work-function of stable $\rm  CsK_2Sb$ surfaces, combined with the high absorption coefficient in the visible and ultraviolet regions, suggests that this material is well-suited for generating high-brightness electron beams in linear accelerators.

Table \ref{tab:table2} further reveal that the Sb-terminated (111)-Sb surface in the unrelaxed slab, which undergoes significant surface reconstruction to stabilize into a K-terminated structure, has the highest work-function, around 3.29 eV. However, due to its high surface energy, the (111)-Sb slab is not the preferred exposed surface during crystal growth. The other metallic surfaces have intermediate work-function values. On the other hand, the semiconducting slab (110) has ionization potential of 2.64 eV.

\section*{4 Conclusions}
In summary, we investigated the electronic, elastic, vibrational, and electrical transport properties of $\rm CsK_2Sb$. Our study predicts that $\rm CsK_2Sb$ has a bandgap of 1.13 eV. Unlike $\rm K_3Sb$ and $\rm Cs_3Sb$, where the distorted phase is more stable, ordered $\rm CsK_2Sb$ is found to be structurally stable. At room temperature, the electron and hole mobility of $\rm CsK_2Sb$ are predicted to be 111.86 $\rm cm^2/Vs$ and 3.24 $\rm cm^2/Vs$, respectively. 

Our findings indicate that vacancy defects are the dominant defects in $\rm CsK_2Sb$. Defects in $\rm CsK_2Sb$ introduce electronic states within the bandgap, leading to low-energy peaks in the absorption spectrum. The calculated absorption coefficient varies by approximately $\pm 4\%$ at 2.5 eV compared to $\pm 10\%$ at 3.6 eV, suggesting that defects can significantly influence photoemission in $\rm CsK_2Sb$.  
Our analysis indicates that vacancy clusters formed by the removal of a cation and an anion exhibit strong binding, despite their relatively high formation energy compared to di-vacancies consisting of two Cs or K atoms.

Our surface property analysis reveals that the stability of the $\rm CsK_2Sb$ surface depends on the chemical potentials of Cs and K and that tuning the environment during growth conditions one should be able to control the type of surface termination. Under Cs-rich conditions, the (111)-Cs surface remains stable regardless of $\mu_{\rm K}$. For intermediate $\mu_{\rm Cs}$, the stability shifts among the (110), (001)-CsSb, and (001)-K terminated surfaces, depending on $\mu_{\rm K}$. As the conditions transition toward Cs-poor environments, the (111)-K(b) terminated surface becomes stable over a broad range of $\mu_{\rm K}$.  
The (111)-Cs-terminated surface exhibits the lowest work-function of 2.14 eV, which is significantly lower than that of metallic photocathodes. This small work-function suggests that visible light can efficiently induce electron emission in $\rm CsK_2Sb$.

Though $\rm CsK_2Sb$ demonstrates highly promising electronic, optical, transport, and surface properties for photocathode applications, its high cesium/potassium content makes it extremely reactive toward residual gases, posing challenges for long-term stability and requiring stringent vacuum or encapsulation measures to approach the operational lifetimes of metallic photocathodes.\cite{G17_2017, WYNB2018, vitaly_2016, GLYHYM_2020} Additionally, our surface stability analysis highlights the delicate dependence of surface terminations on growth conditions, emphasizing the need for precise control of chemical potentials during synthesis. The presence and nature of defects, especially vacancy clusters, also play a significant role in modifying the photoemission behavior. The calculations presented here could help experimental
group select the range of conditions to control the growth of the best crystal, consequently optimizing $\rm CsK_2Sb$ for high-performance photocathode applications.

\section*{Acknowledgements}
The authors acknowledge the Laboratory Directed Research and Development program of Los Alamos National Laboratory (LANL) for funding this project. Los Alamos National Laboratory is operated by Triad National Security, LLC, for the National Nuclear Security Administration of the U.S. Department of Energy (contract no. 89233218CNA000001). We used  LANL’s institutional supercomputer to perform the calculations for this project.

\bibliography{refs}

\end{document}